%% file: acmmanuscript.tex
\documentclass[manuscript,screen,acmsmall,nonacm]{acmart}
\AtBeginDocument{%
  }

\usepackage{graphicx} % Required for inserting images
\usepackage{algorithm}
\usepackage{algpseudocode}
\usepackage{xcolor}
\usepackage{amsmath}
\usepackage{amsfonts}
\usepackage{bbm}
\usepackage{booktabs}
\usepackage{hyperref}
\usepackage{tikz}
\usepackage{comment}
\usetikzlibrary{mindmap,trees,calc,arrows.meta, bending}
\usepackage{dirtytalk}

\definecolor{CentralColor}{HTML}{DDDDDD}
\definecolor{HexagonOne}{HTML}{2A9D8F}
\definecolor{HexagonTwo}{HTML}{8AB17D}
\definecolor{HexagonThree}{HTML}{E9C46A}
\definecolor{HexagonFour}{HTML}{F4A261}
\definecolor{HexagonFive}{HTML}{E76F51}
\definecolor{HexagonSix}{HTML}{9B5DE5}

\def\HexRadius{1.75}

\newcommand{\Hexagon}[3]{%
    \draw[
        fill=#2,
        fill opacity=0.45, % Hexagon transparency
        draw=none         % Remove borders
    ]
    ($(#1)+(30:\HexRadius)$)
    \foreach \angle in {90,150,210,270,330} {
        -- ($(#1)+(\angle:\HexRadius)$)
    }
    -- cycle;

    \node[
        align=center,
        text=black,       % Black text
        font=\sffamily
    ] at (#1) {#3};
}

\newcommand{\HexEdgeLabel}[3]{%
    \ifcase#2\relax
        \PackageError{HexEdgeLabel}{Edge number must be between 1 and 6}{}%
    \or
        \def\LabelAngle{180}\def\LabelRotation{90}%  Edge 1: left
    \or
        \def\LabelAngle{120}\def\LabelRotation{30}%  Edge 2: upper left
    \or
        \def\LabelAngle{60}\def\LabelRotation{-30}%  Edge 3: upper right
    \or
        \def\LabelAngle{0}\def\LabelRotation{270}%    Edge 4: right
    \or
        \def\LabelAngle{-60}\def\LabelRotation{30}%  Edge 5: lower right
    \or
        \def\LabelAngle{-120}\def\LabelRotation{-30}% Edge 6: lower left
    \else
        \PackageError{HexEdgeLabel}{Edge number must be between 1 and 6}{}%
    \fi
    \node[
        rotate=\LabelRotation,
        text=black,
        font=\sffamily\scriptsize,
        align=center
    ] at ($(#1)+(\LabelAngle:\HexRadius+.15)$) {#3};
}

\acmDOI{}
\renewcommand{\keywordsname}{\textbf{Keywords}}
\begin{document}

\title{Neurosymbolic Human–AI Collaboration: Foundations, Conceptual Framework, and Applications}

\author{Vito Scaraggi}
\email{vito.scaraggi@iit.cnr.it}
\orcid{0009-0002-9243-0309}
\affiliation{%
  \institution{Institute of Informatics and Telematics (IIT), University of Pisa}
  \city{Pisa}
  \country{Italy}
}

\author{Andrea Passarella}
\email{andrea.passarella@iit.cnr.it}
\author{Marco Conti}
\email{marco.conti@iit.cnr.it}
\affiliation{%
  \institution{Institute of Informatics and Telematics (IIT)}
  \city{Pisa}
  \country{Italy}
}

\renewcommand{\shortauthors}{Scaraggi et al.}

\begin{abstract}
  The growing adoption of Artificial Intelligence (AI) across workplaces and industries, ranging from manufacturing plants to healthcare facilities, has paved the way to the design of systems in which humans and AI collaborate on shared tasks. The first developments of Human–AI Collaboration (HAIC) were driven by advances in Machine Learning (ML), which enabled the automation of numerous tasks, including large-scale data processing, image analysis, and speech recognition. However, ML automation have proven insufficient for certain applications due to methodological, safety, and ethical concerns. In such cases, researchers argue for integrating human expertise into operational workflows to improve performance, relying on the complementary skill sets of humans and AI. Despite long-standing research, the field of HAIC has not yet reached full maturity, largely due to the intrinsic difficulty of ensuring a seamless communication between humans and AI. This challenge is partially addressed by Explainable AI methods, which aim to make internal model reasoning more transparent through human-understandable explanations. Recognizing the need for a shared interface between humans and AI in collaborative settings, we propose a novel conceptual framework, NeSy-HAIC, that leverages Neurosymbolic AI (NeSy AI) to bridge the semantic gap between machine representations and human-understandable symbolic concepts. The framework builds on current research in NeSy AI and HAIC, supports multiple interaction paradigms, including Human-In-The-Loop and Human-On-The-Loop, and is characterized through its enabling methodologies and two illustrative use cases in semi-autonomous driving and security alert prioritization. We argue that the integration of data-driven learning and symbolic reasoning can substantially enhance human–AI interaction and mitigate limitations of collaborative systems.
\end{abstract}

\begin{comment}
\begin{CCSXML} %todo
<ccs2012>
 <concept>
  <concept_id>00000000.0000000.0000000</concept_id>
  <concept_desc>Do Not Use This Code, Generate the Correct Terms for Your Paper</concept_desc>
  <concept_significance>500</concept_significance>
 </concept>
 <concept>
  <concept_id>00000000.00000000.00000000</concept_id>
  <concept_desc>Do Not Use This Code, Generate the Correct Terms for Your Paper</concept_desc>
  <concept_significance>300</concept_significance>
 </concept>
 <concept>
  <concept_id>00000000.00000000.00000000</concept_id>
  <concept_desc>Do Not Use This Code, Generate the Correct Terms for Your Paper</concept_desc>
  <concept_significance>100</concept_significance>
 </concept>
 <concept>
  <concept_id>00000000.00000000.00000000</concept_id>
  <concept_desc>Do Not Use This Code, Generate the Correct Terms for Your Paper</concept_desc>
  <concept_significance>100</concept_significance>
 </concept>
</ccs2012>
\end{CCSXML}

\ccsdesc[500]{Do Not Use This Code~Generate the Correct Terms for Your Paper}
\ccsdesc[300]{Do Not Use This Code~Generate the Correct Terms for Your Paper}
\ccsdesc{Do Not Use This Code~Generate the Correct Terms for Your Paper}
\ccsdesc[100]{Do Not Use This Code~Generate the Correct Terms for Your Paper}
\end{comment}



\settopmatter{printacmref=false}
\setcopyright{none}
\maketitle

\input{section_intro_rel}
\input{section_nesy}
\input{section_haic}
\input{section_framework}
\input{section_conclusion}

%\begin{acks}
%To Robert, for the bagels and explaining CMYK and color spaces.
%\end{acks}

\begin{comment}
\section*{Ethics and Privacy Statement}

This section of your ACM work should discuss the potential societal
risks that might result from its publication; two to three sentences
related to the findings of your study, or new advancements made
possible by their developed methods. The privacy and ethics statement
should clearly address the broader impacts of their work as it relates
to the authors' interpretation of privacy, fairness, safety, human
rights, data sovereignty, or future misuse and any benefit/risk
trade-off resulting from this research. We acknowledge that some
papers may have minimal societal risks beyond those considered by
institutional review boards, and the dimensions considered by any
review of the user study design or dataset licenses could be provided
in this statement.
\end{comment}

\bibliographystyle{unsrtnat}
\bibliography{bibliography}

\end{document}

%% file: section_intro_rel.tex
\section{Introduction} \label{sec:intro}
The widespread use of Artificial Intelligence (AI) across many industries has heavily impacted decision-making workflows and increased productivity by automating manual tasks and augmenting human skills. Despite the potential for full automation, ML-based techniques often face difficulties related to data availability, algorithmic biases, and alignment with human expectations. These limitations have motivated research toward integrating human expertise into decision-making models to improve not only predictive performance but also safety and ethical compliance \cite{Kumar2024}. Human-AI Collaboration (HAIC) proposes to convey complementary strengths of AI and humans in a synergistic framework to support a wide range of tasks, spanning domains from manufacturing to healthcare \cite{Khosravy2024, Gao2024}. In fact, while AI is generally most valuable for speed, scale, consistency, information retrieval, and high-confidence pattern recognition, humans excel in handling contextual knowledge, difficult cases, accountability, and deciding when an AI output should be accepted or overridden. These complementary capabilities can enhance system robustness and mitigate failures arising from individual weaknesses.

Although HAIC has led to significant improvements in many application domains, a large body of work suggests that integrating human expertise with AI does not necessarily achieve better results than either the human or the AI alone. Analyzing a corpus of 370 HAIC-based experiments, Vaccaro et al. \cite{Vaccaro2024} found that, on average, human-AI systems performed worse than at least one of the human alone or the AI alone, demonstrating how collaborative settings can potentially hinder efficiency. The study revealed that human-AI teams are more likely to experience performance losses in decision tasks rather than creative tasks, and also when the AI alone outperforms the human alone. In general, challenges like trust issues, e.g., overreliance and underreliance on AI outputs, and communication barriers often undermine the productivity of collaboration between humans and AI. To overcome these problems, HAIC systems must be carefully engineered to support seamless interaction between humans and AI. The major design choices include \textit{what} AI should communicate to the human, \textit{when} the interaction should occur, and \textit{how} the human should interface with the AI \cite{Gomez2024}. These human-centered aspects of collaboration are often overlooked in favor of enhancing AI capabilities, thus degrading the effectiveness of the collaboration.

Current research offers multiple pathways towards improving HAIC design and mitigating or solving its latent issues. A straightforward approach is to seek human-AI complementarity through a proper task delegation strategy. Assuming that humans are better at some subtasks and the AI is better at others, a carefully designed HAIC system should allocate subtasks to whichever team member is the best for that subtask \cite{Donahue2022}. Assigning parts of the task to the partner best suited to perform them can foster effective specialization and improve average team performance. Another crucial aspect of collaboration lies in the capability of an AI system to communicate the rationale behind its predictions and provide contextual information to the human overseeing the decision-making process. Explainable AI (XAI) investigates methods to accomplish interpretability and transparency within AI models, through techniques such as model visualization, feature attribution, and natural language explanations \cite{Kumar2024}. XAI can produce human-readable justifications for AI outputs, giving insights into the internal model's reasoning and facilitating human downstream operations \cite{Senoner2024}. The explanation generated is all the more useful the more it can be correctly interpreted by users and help them in decision-making. 

Research interest in XAI is largely driven by the need to establish an effective communication interface between humans and AI systems to support smooth interaction between the two. Even when humans and AI perform well individually, real collaboration requires a shared interaction substrate that enables humans to understand AI behavior and, conversely, AI systems to interpret human intentions and needs. The authors in \cite{Kambhampati2022} argue that this common interface has to be composed of \textit{symbols}, representing a vocabulary of explicit concepts expressed in a formal language known to humans. To enable symbols to serve as a \textit{lingua franca} for human–AI interaction, they must be appropriately encoded into the representations on which AI models can operate. Research exploring AI models that can understand symbols finds a natural foundation in Neurosymbolic AI (NeSy AI), which investigates approaches merging symbolic representations with common machine learning algorithms, such as neural networks. Neurosymbolic methods are promising candidates for bridging the gap between human cognition and AI in collaborative scenarios, as they combine the interpretability of symbolic reasoning with the representational and learning capabilities of data-driven methods. Despite the fitness of NeSy AI for HAIC, collaborative frameworks explicitly leveraging neurosymbolic integration are still underexplored.

Acknowledging this research gap, in this work we propose a Neurosymbolic Human-AI Collaboration (NeSy-HAIC) framework that accommodates multiple interaction modes between humans and AI and deeply integrates NeSy AI as an essential building block of the collaboration process. The proposed architecture adapts to different levels of cooperation and task types, and its applicability is demonstrated through real-world use cases that instantiate NeSy-HAIC. Through the proposed framework, we aim to provide a blueprint for modeling future HAIC tasks while consolidating prior research efforts into a unified conceptual model. To contextualize our proposal in the current literature, we first discuss the main foundations of NeSy AI and HAIC. In Section \ref{sec:rel} we survey several related works that refer either to NeSy AI or HAIC. In Section \ref{sec:nesy} we briefly present the fundamentals of NeSy AI. In Section \ref{sec:haic} we introduce HAIC through a dimensional analysis and illustrate HAIC applications in three selected domains, namely healthcare, cybersecurity, and robotics. Furthermore, we motivate the adoption of NeSy AI in HAIC contexts, given the limitations of HAIC methods currently used in the analyzed application domains, and review a limited number of studies that address neurosymbolic integration in collaborative settings. In Section \ref{sec:framework}, we describe the NeSy-HAIC framework, its internal components and features, and the enabling methodologies that support its practical instantiation. We also present two use cases of the framework in semi-autonomous driving and alert prioritization tasks, illustrating how NeSy-HAIC can be customized to fit their specific requirements and comparing the resulting systems with existing HAIC solutions. Finally, in Section \ref{sec:concl}, we discuss challenges and limitations of our proposal.
Figure \ref{fig:survey_structure} illustrates the organizational structure of the paper.

\input{img/survey_structure}

\section{Related work} \label{sec:rel}
The idea of leveraging NeSy AI to foster HAIC is relatively novel, with only a limited number of studies hinting at this possibility. Nawaz et al. \cite{Nawaz2025} presented the state-of-the-art of NeSy AI and analyzed its contribution to advanced cognitive systems that incorporate data-driven capabilities, high-level reasoning, and interpretability. After a historical overview, the survey delves into the core components of NeSy AI, namely symbolic reasoning, referred to as logic manipulation of symbols and knowledge bases, and the different neural learning mechanisms. Then the authors described notable neurosymbolic algorithms, such as Logic Tensor Networks (LTN), which convert First-Order Logic (FOL) statements into real-valued tensors and adopt a satisfiability loss for rule enforcement, and DeepProbLog, which combines probabilistic reasoning with neural networks. These methods are compared in terms of accuracy, scalability, interpretability, and computational efficiency and evaluated through a SWOT analysis. Subsequently, the authors discussed applications of Neurosymbolic AI, e.g., robotics and healthcare. Although the paper does not directly address the use of NeSy AI in collaborative settings, it highlights the potential contribution of NeSy AI to the development of advanced ethically aligned systems capable of interacting with humans in sectors like healthcare and finance.

Miedema et al. \cite{Miedema2026} analyzed three main topics: decision-making architectures that involve humans and AI systems, the trustworthiness of AI systems, and knowledge- and data-driven AI (i.e., NeSy AI). The authors differentiated decision-making systems into autonomous systems that operate without human intervention, supportive systems aiding human decision-makers, and collaborative systems building on top of complex interaction dynamics between humans and AI. The work highlights the distinction between interpretability, that is, the measure of how easily a human can understand AI models' behavior, and explainability, that is, the capability to justify model outputs. While XAI can guarantee the second property, NeSy AI is needed to build actually interpretable algorithms. The paper covers the fundamentals of NeSy AI, including the Kautz taxonomy, knowledge representation, reasoning, learning, and evaluation techniques, then presents a lifecycle framework for knowledge- and data-driven AI systems, articulated through six phases and mapped onto a healthcare example. Note that the proposed lifecycle model fundamentally differs from our NeSy-HAIC framework, as it describes the operational stages from system design to deployment and maintenance at a high level, whereas our framework focuses more specifically on human–AI interaction and neurosymbolic integration.

Kumar et al. \cite{Kumar2024} surveyed the state of the art in HAIC applications, which they refer to as Human-in-the-Loop (HITL) systems. The major contributions span the description of main methodologies to incorporate human feedback into ML (e.g., Active Learning), the discussion of challenges associated with HITL systems, and the illustration of case studies in specific domains, such as autonomous driving and healthcare. Although the study does not explicitly discuss the role of NeSy AI in HAIC, it clearly identifies several recurring challenges in collaborative systems, many of which (e.g., interpretability and ethical issues) could potentially be mitigated through neurosymbolic integration. Therefore, this work provided valuable insights for our study.

Gomez et al. \cite{Gomez2024} extracted a taxonomy of human-AI interaction patterns from a corpus of more than 100 studies and analyzed their frequency in different application domains. This work provides a systematic overview of approaches to HAIC and offers a detailed characterization of how human and AI contributions are interleaved within collaborative tasks. The authors distinguished between AI-first assistance and AI-follow assistance, depending on whether the AI acts before or after human intervention, and described several interaction patterns through which humans and AI can engage, including human-initiated assistance requests, AI-initiated dialogue, and explicit task delegation mechanisms. Among the identified works, AI-first and AI-follow assistance are the most prevalent interaction patterns. However, these patterns are not mutually exclusive, and some studies combine multiple forms of interaction.

Kambhampati et al. \cite{Kambhampati2022} advocated for a novel paradigm for enhancing human-AI interaction based on a symbolic interface that supports human-understandable concepts. The proposal is motivated by the need to equip AI models with the ability to provide explanations to humans and incorporate human advice, using a shared communication layer. As mentioned in the paper, the development of such a symbolic interface involves three main phases: first, the construction of an initial vocabulary of concepts related to the task domain and the AI capabilities seen from a human perspective; second, the grounding of these symbolic concepts into representations that can be processed by AI models; and finally, the progressive expansion and refinement of the concept vocabulary through human feedback to ensure that the symbolic interface remains consistent over time. The conceptual framework is illustrated through the examples of real systems providing symbolic explanations to humans or interpreting human advice. Overall, this work is motivated by similar considerations to ours and similarly recognizes the importance of neurosymbolic approaches for HAIC.

%% file: img/survey_structure.tex
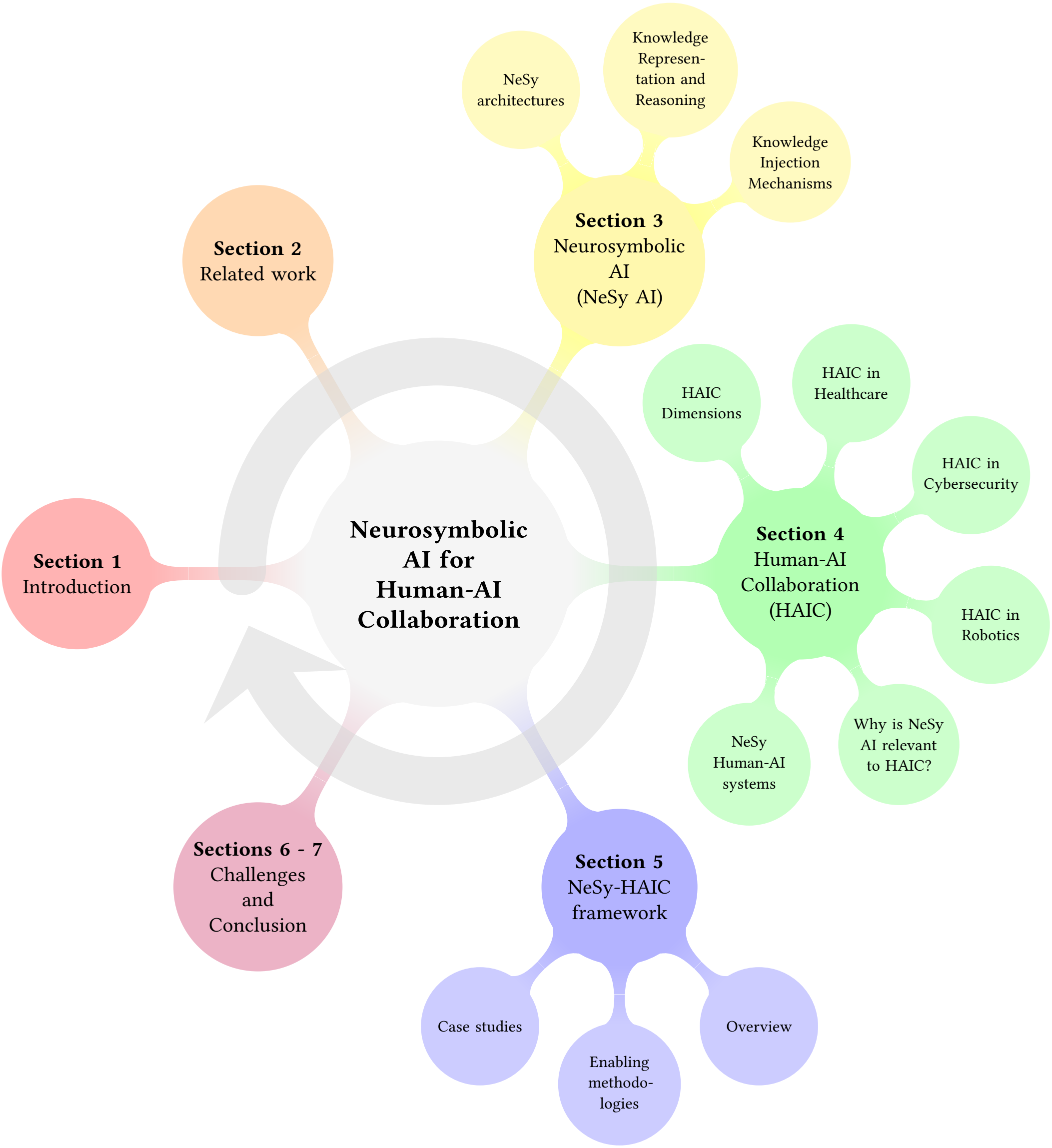
\begin{figure}[ht!]
\centering
\begin{tikzpicture}[scale=.87, transform shape, mindmap, grow cyclic, every node/.style=concept, concept color=black!4, 
	level 1/.append style={level distance=5.5cm,sibling angle=60, clockwise from=180},
	level 2/.append style={level distance=3cm,sibling angle=45, font=\fontsize{7pt}{9pt}\selectfont}]
\node{\textbf{Neurosymbolic AI for \\ Human-AI Collaboration}}
child[concept color=red!30]{ node{\textbf{Section 1} %\ref{sec:intro} 
\\ Introduction }
}
child[concept color=orange!30, level 2/.append style={sibling angle=40}]{ node{\textbf{Section 2} %\ref{sec:rel} 
\\ Related work }
}
child[concept color=yellow!40] { node{\textbf{Section 3} %\ref{sec:nesy} 
\\ Neurosymbolic AI \\ (NeSy AI)} [clockwise from=120]
    child[concept color=yellow!30] { node{NeSy \\ architectures} }
    child[concept color=yellow!30] { node{Knowledge Representation and Reasoning} 
    }
    child[concept color=yellow!30] { node{Knowledge Injection Mechanisms} }
}
child[concept color=green!30]{ node{\textbf{Section 4} %\ref{sec:haic} 
\\ Human-AI \\ Collaboration \\ (HAIC)} [clockwise from=120]
    child[concept color=green!20] { node{HAIC Dimensions} }
    child[concept color=green!20] { node{HAIC in Healthcare}
    }
    child[concept color=green!20] { node{HAIC in Cybersecurity} }
    child[concept color=green!20] { node{HAIC in Robotics} }
    child[concept color=green!20] { node{Why is NeSy AI relevant to HAIC?} }
    child[concept color=green!20] { node{NeSy \\ Human-AI systems} }
}
child[concept color=blue!30]{ node{\textbf{Section 5} %\ref{sec:framework} 
\\ NeSy-HAIC framework} [clockwise from=-45]
    child[concept color=blue!20] { node{Overview} }
    child[concept color=blue!20] { node{Enabling \\ methodo- logies} }
    child[concept color=blue!20] { node{Case studies} }
}
child[concept color=purple!30]{ node{\textbf{Sections 6} %\ref{sec:concl} 
\textbf{- 7} %\ref{sec:concl2} 
\\ Challenges and \\ Conclusion}
}
;

% Adjustable arrow parameters
\newcommand{\ArrowXShift}{-38mm}       % positive: right; negative: left
\newcommand{\ArrowYShift}{-31mm}       % positive: up; negative: down
\newcommand{\ArrowLineWidth}{18pt}  % thickness of the curved arrow

\begin{scope}[
    opacity=0.55,
    transparency group
]

% Clockwise arrow surrounding the mind map
\draw[
    -{Stealth[length=10mm,width=20mm]},
    line width=\ArrowLineWidth,
    line cap=round,
    line join=round,
    draw=black!15
]
let
    \p1=(current bounding box.south west),
    \p2=(current bounding box.north east),
    \p3=($(\p1)!0.5!(\p2)$),
    \p4=($(\p3)+(\ArrowXShift,\ArrowYShift)$),
    \n1={0.5*(\x2-\x1)-50mm},
    \n2={0.5*(\y2-\y1)-55mm}
in
    ($(\p4)+({\n1*cos(100)},{\n2*sin(100)})$)
    arc[
        start angle=180,
        end angle=-165,
        x radius=\n1,
        y radius=\n2
    ];
\end{scope}

\end{tikzpicture}
\caption{Mind map representing the organizational structure of the paper}
\label{fig:survey_structure}
\end{figure}

%% file: section_nesy.tex
\section{Neurosymbolic AI} \label{sec:nesy}

Neurosymbolic AI (NeSy AI) is a family of AI methods that combine symbolic and subsymbolic techniques to jointly learn from data and perform logical reasoning \cite{Bhuyan2024}. The \say{neuro} component refers to data-driven approaches, such as neural networks, that can generalize from partial and noisy unstructured information; the \say{symbolic} part conveys knowledge in the form of a predefined set of symbols and rules and reasons over these representations. The integration of the two paradigms, which finds a notable parallel in Kahneman's theory of human cognition \cite{Kaheman2011}, seeks to exploit their complementary strengths while mitigating their respective limitations. Symbolic AI enables sound human-like reasoning; however, it is severely constrained by its \say{brittleness} when information is incomplete and computational intractability in complex environments and open-world settings. Conversely, subsymbolic AI excels at learning from data, but often lacks explicit reasoning capabilities and interpretability. The range of strategies available for neurosymbolic integration gives rise to systems with diverse architectural and functional properties. In the following sections, we systematize this space by examining its main design choices.

\subsection{Neurosymbolic architectures}
A notable classification of neurosymbolic architectures is presented in \cite{Kautz2022}, which delineates six principal categories organized in ascending order of maturity with respect to the type of neurosymbolic integration. This scale spans from \textit{Type 1, or Symbolic Neuro-Symbolic} systems, which correspond to conventional neural networks, to \textit{Type 6, or Neuro[Symbolic]} systems, which represent the theoretical ideal of fully integrated symbolic and subsymbolic components, even if the author claimed that no existing framework realized this level of integration. The in-between classes include \textit{Type 2, or Symbolic [Neuro]}, which adopts a neural subroutine as a heuristic function within a symbolic problem solver, as DeepMind's AlphaGo \cite{Silver2016}, \textit{Type 3, or Neuro|Symbolic}, which places symbolic and subsymbolic components in parallel as the Neurosymbolic Visual Question Answering (NS-VQA) \cite{Yi2018}, \textit{Type 4, or Neuro: Symbolic $\rightarrow$ Neuro}, which uses a neural network with a specific training regime based on symbolic rules \cite{Lample2019}, and \textit{Type 5, or Neuro\_Symbolic}, which encodes symbolic rules into loss function and network weights, enabling gradient-based learning with knowledge, as in Logic Tensor Networks \cite{Badreddine2022}. More recently, several taxonomies based on different classification criteria have been proposed \cite{vanBekkum2021, Hamilton2024, Amador2024}, underscoring the absence of a single, universally accepted taxonomy for neurosymbolic systems. 

\subsection{Knowledge Representation and Reasoning}
Neurosymbolic systems exhibit great variation in how knowledge is represented and manipulated for reasoning. Logic-based languages can represent both rules and facts, with different formalisms offering varying levels of expressiveness and guarantees regarding formal properties.
Propositional Logic (PL) can formulate boolean statements, built by combining multiple propositions using logical operators that can be evaluated as \textit{false} (0) or \textit{true} (1). First-Order Logic (FOL) extends PL by adding constants, variables, predicates, and quantifiers, allowing the representation of more complex concepts. In contrast to classical logic languages such as PL and FOL, which are limited by two-valued semantics, Fuzzy Logic (FL) is a form of multi-valued logic wherein statements can take on any truth value within the continuous interval $[0,1]$ (\,$[$\textit{false, true}$]$\,). To render FL computable, it is necessary to use numerical functions that approximate logical operators, as traditional truth tables used in classical logic are incompatible with continuous representations. Although FL does not preserve all properties of classical logic, it facilitates a more seamless integration with subsymbolic approaches, enabling differentiable end-to-end training. Other possible choices are Linear Temporal Logic (LTL) \cite{Sun2024}, which is used to reason about the temporal properties of systems; Answer Set Programming (ASP), which is a form of declarative programming to formulate complex problems in terms of rules and constraints; and Domain-Specific Languages (DSLs), which are declarative programming languages tailored to specific application domains.

Other efficient tools for representing knowledge are based on graph structures. Knowledge Graphs (KG) represent domain-specific objects and their semantic relationships as a set of triples $G=(h,r,t)$, where $h$ and $t$ are entities connected by a relation $r$, and are described by schema languages such as RDF or OWL. In classification tasks, Decision Trees (DT) arrange decision-making routes in a tree-like structure, in which internal nodes represent an attribute-based decision split and leaf nodes contain the final classification output. For domains characterized by uncertainty, probabilistic graphical models, such as Bayesian networks and Markov networks, provide apt graph representations to compactly encode complex probability distributions over high-dimensional variable spaces \cite{Koller2009}.

Beyond knowledge representation, a second fundamental component of a symbolic system is the reasoning algorithm used to derive new knowledge from existing information. For logic languages, the term \say{reasoning} may refer to various forms of inference, such as \textit{deductive reasoning}, in which conclusions are drawn from general rules and specific instances, \textit{abductive reasoning}, in which the best explanation for a set of observations is inferred, or \textit{inductive reasoning}, in which generalizations are made based on specific examples. Reasoning over knowledge graphs generally relies on graph-based inference mechanisms, including path traversal and multi-hop reasoning, which obtain implicit relationships by following sequences of connected entities \cite{Lin2018}. In contrast, reasoning in probabilistic graphical models involves computing or approximating probability distributions and posterior beliefs, typically through inference algorithms such as variable elimination, belief propagation, or sampling-based methods.

\subsection{Knowledge Injection Mechanisms} \label{sec:sgp}
A long-standing open challenge in NeSy AI is the \textit{Symbol Grounding Problem} \cite{Harnad1990}. This problem concerns the fundamental challenge of bridging the gap between abstract symbolic representations and neural networks, as the latter can only process numerical data. Numerous Symbolic Knowledge Injection (SKI) mechanisms have been proposed to translate symbolic knowledge into a representation compatible with subsymbolic methods \cite{Agiollo2023}. \textit{Loss Constraining} is a technique that increases the loss function with penalty terms whenever some logical formulae are violated during network training\cite{Badreddine2022}. In contrast, \textit{Structural Constraining} shapes all or part of the neural network architecture to reflect the evaluation process of one or more logical formulae \cite{Riegel2020}, while \textit{Embedding} methods transform knowledge into real-valued tensors used for training, as in Knowledge Graph Embedding (KGE) approaches. Additional types of neurosymbolic integration belong to the field of Neurosymbolic Reinforcement Learning \cite{Acharya2024}. Some works \cite{Garnelo2016, Garcez2018, Mitchener2022} proposed a neural backend capable of extracting meaningful representations for further symbolic reasoning, while others adopted symbolic methods for supporting reward shaping \cite{Velasquez2021}, representing an agent's policy as a short DSL program \cite{Verna2019}, or initializing the RL policy through a warmup process guided by human expert knowledge \cite{Silva2021}.

%% file: section_haic.tex
\section{Human-AI Collaboration} \label{sec:haic}
In general, Human-AI Collaboration (HAIC) is the process through which humans and AI collaborate to achieve a common goal \cite{Puerta-Beldarrain2025}. In HAIC applications, the team is composed of at least one human and one AI agent, often having predefined roles and communicating with specific protocols. The concept of HAIC spans multiple research fields, each of which often adopts its own formulation and terminology. Although a unified definition of HAIC is still lacking, different research domains agree that HAIC systems should exploit the complementary strengths of humans and AI to improve overall team performance. In fact, humans are best suited for complex reasoning assignments that require commonsense and domain knowledge, while AI is essential in data-driven and computation-heavy tasks. An effective HAIC strategy also seeks to compensate for the limitations of each party, including human fatigue and the susceptibility of AI systems to algorithmic biases and hallucinations. 

\subsection{HAIC Dimensions}
Designing an HAIC system can be challenging because of the inherent complexity of integrating human and AI agents. Hence, we discuss a set of design dimensions for developing such systems, as illustrated in Figure \ref{fig:haic_dim}. These reflect some design principles outlined in \cite{Puerta-Beldarrain2025, Mostafa2019, Punzi2026, Mosqueira-Reya2023} and encompass multiple key concepts such as the role of humans within hybrid teams, the degree of autonomy, and adaptation. \\

\input{img/hexagons}

\textit{Goal of collaboration.} HAIC can pursue different objectives depending on the application context. In some cases, AI is employed to augment human intelligence by extending human analytical, decision-making, or creative capabilities. In others, human involvement is intended to supervise AI decisions for which interpretability and explainability are essential, particularly in settings where users must understand, evaluate, or contest system outputs. Collaboration may also enhance system robustness by combining complementary human and machine capabilities, while the delegation of repetitive or cognitively demanding activities to AI can reduce human workload and fatigue. Thus, HAIC should not be understood as serving a single purpose, but rather as encompassing a range of configurations designed to improve human capabilities, system trustworthiness, and overall joint performance \cite{Wang2026}. \\

\textit{Human role.} Humans can assume diverse roles within hybrid teams depending on their degree of expertise and their use of the system. They can be simple end-users without any prior domain knowledge, domain experts who oversee the training, or expert users who cooperate with AI systems when these are deployed. In active learning, humans act as experts who provide annotations for unlabeled data \cite{Mosqueira-Reya2023}. In this process, AI models actively ask humans for guidance to deal with uncertain input instances and use their feedback to refine predictions. Alternative methods treat humans as teachers \cite{Simard2017}, giving them complete control over the learning process. The teacher can construct an ordered representation of knowledge, commonly referred to as \textit{curriculum}, that is used to train an AI model with increasingly complex tasks. Beyond providing feedback or teaching models, humans can also work in parallel with AI. In this setting, one of the main challenges lies in effectively switching control from humans to AI agents, a task known as delegation \cite{Fuchs2023}. Delegation mechanisms should determine when to defer and ensure a smooth transition between agents in control, pursuing enhanced team performance and robustness to individual failures. \\

\textit{Architecture.} From an architectural perspective, HAIC systems can be classified into Human-In-The-Loop (HITL) and Human-On-The-Loop (HOTL). HITL is characterized by the human actively participating in most, if not all, control decisions and system iterations, while in HOTL the human assumes a supervisory role and takes over control only when system failures or unexpected events occur \cite{Nothwang2016}. HITL guarantees stricter control over the whole decision-making process, which is mandatory in safety-critical contexts. In contrast, HOTL is preferred when having humans constantly in charge of control is unfeasible or too demanding. Architectural solutions for HAIC may differ according to the stage at which human intervention occurs. For instance, humans can intervene after the AI model makes a prediction by providing corrections and feedback to the system, or before the AI model, by filtering training data or supplying additional input. Human and AI agents can work sequentially, if one agent processes the output of another agent, or in parallel, when both agents operate independently at the same time \cite{Hekler2019}. One particularly important design consideration is the team composition. A team may comprise a single human and a single AI agent or include multiple specialized AI agents, as explored in Mixture-of-Experts (MoE) approaches. \\

\textit{Autonomy.} Autonomy is the faculty of an AI agent to execute a task independently. The degree of AI autonomy should not be regarded as a static attribute, but rather as a continuum from full human control to full AI autonomy, along which an HAIC system may dynamically reposition itself in response to changing environmental conditions. Many approaches have been proposed to implement Adjustable Autonomy, such as predictive systems switching autonomy levels based on expected performance \cite{Bush2012} or policy-based systems allowing autonomous agents to adjust their behavior autonomously \cite{Alzahrani2013}. \\

\textit{Adaptation.} The effectiveness of HAIC largely depends on the ability of AI models to adapt to human behavior and shifts in environmental conditions. An adaptive model should understand human intentions, detect changes, and modify itself accordingly to maintain the quality of collaboration. To achieve adaptation, AI agents leverage methods such as learning from feedback, human demonstrations, and interactive collaboration to refine their behavior over time. Central to the concept of adaptation is the development of a \say{world understanding} within AI models, in the form of a mental model that captures goals, other agents' intentions, and environment mechanics. Mental models allow agents to anticipate other agents' behavior and act accordingly, ensuring situation awareness \cite{Endsley2023} and smooth coordination. \\

\textit{Enabling methodologies.} The implementation of HAIC frameworks requires specific techniques designed for enabling human-AI interaction. Interactive Machine Learning (IML) \cite{Fails2003} denotes a class of methods in which humans can examine the intermediate outcomes of a model, provide feedback, and intervene directly in the training process. Integrating a human into the loop may involve allowing them to select informative samples for model training or to correct model outputs to refine subsequent predictions. IML also incorporates Human–Computer Interaction (HCI) methods aimed at facilitating communication between humans and AI models, including user-friendly graphical interfaces and collaborative systems that account for human factors. Active Learning (AL) involves an iterative process in which models query human annotators to label selected data points and subsequently use the acquired labels for training. The procedure generally relies on a sampling strategy that prioritizes outliers or samples associated with uncertain predictions, and continues until a predefined performance threshold is reached \cite{Mosqueira-Reya2023}. Although AL is widely used to maximize accuracy while minimizing the number of human-labeled samples, it often relies on the unrealistic assumption of an error-free human oracle, and its iterative nature can incur substantial computational costs due to repeated model training. Reinforcement Learning (RL) is commonly used to build autonomous agents capable of interacting with an environment and learning a behavioral policy based on a reward function. When the environment includes human agents, the learned policy may emerge from collaborative interactions or be explicitly aligned with human values, as in Reinforcement Learning from Human Feedback (RLHF) \cite{Kaufmann2025}, where the agent optimizes a reward model learned from preferences provided by a human-in-the-loop rather than relying on a predefined reward function. Alignment with human preferences can be also achieved by learning from demonstrations provided by humans, as in Imitation Learning (IL) and Inverse Reinforcement Learning (IRL) \cite{Zheng2024}.

\subsection{HAIC in Healthcare}
The adoption of the HAIC paradigm has gradually gained importance in healthcare domains, such as diagnostic imaging and clinical decision support systems, benefiting from the combined strength of machine learning and healthcare professionals' knowledge. The hybrid approach complements AI capability to process vast clinical datasets and generate predictions in a timely fashion, with critical decision-making guided by deep expertise and contextual understanding.

A crucial aspect of HAIC systems in healthcare concerns the role assigned to clinicians. Human intervention may be incorporated in various stages of the system lifecycle, for example, during data annotation, model training, or model deployment \cite{Yuan2024}. Depending on the specific HAIC model, experts are committed to verifying AI outputs, or they leverage data-driven insights, such as risk scores and recommended treatments, to formulate an independent judgment \cite{Olawade2026}. Eventually, humans and AI may form a loop where clinicians' feedback is incrementally integrated into machine learning models, as in Active Learning frameworks. Critical and high-risk scenarios are preferably addressed through a human-in-command approach, in which AI serves only as a decision-support tool during the preliminary stages, while the clinician retains final decision-making authority.

Numerous studies revealed that workflows with the joint presence of AI models and healthcare professionals are beneficial from multiple perspectives. Collaborative settings have often demonstrated to yield superior performance compared to full-AI automation and human-only pipelines. For instance, empirical evidence in medical image interpretation revealed that HAIC significantly reduces human review times without compromising diagnostic performance, in cases where AI is used to augment humans, as a second reader or as a pre-screening method. In addition to efficiency, the inclusion of a human-in-the-loop ensures that clinical predictions are aligned with medical standards and transparency measures. This integration also tries to mitigate biases rooted in training data and AI algorithms, which could lead to poor decision-making for an underrepresented portion of the population \cite{Hasanzadeh2025}.

Real-world HAIC systems are deployed in several healthcare sectors. Diagnostic imaging is a popular application where AI algorithms are employed to help clinicians interpret medical images. This use of AI is popular in the radiology field, where diagnostic imaging is applied to chest X‑rays and mammograms to identify potential abnormalities \cite{Sorantin2022}. For example, Patel et al. \cite{Patel2019} introduced an augmentation framework for pneumonia detection in chest radiographs, employing a deep learning model whose output is refined by a swarm intelligence platform composed of radiologists who analyze low-confidence cases in detail. The hybrid strategy achieved higher average detection scores than both the human-only and AI-only approaches used separately, providing justification for adopting HAIC in this setting. Hybrid applications in diagnostic imaging also encompass breast cancer screening, detection of diabetic retinopathy, and classification of skin cancers. 

Clinical Decision Support Systems (CDSS) are capable of handling multiple data modalities, such as medical images, vital signs of the patient, and electronic health records (EHR). Based on these inputs, they produce not only classification results, but also treatment suggestions, risk assessments, and other meaningful outputs that support and enhance clinical decision-making. AI-powered triage systems have become a key technology for improving patient prioritization, reducing average waiting times, and optimizing resource allocations. Timely and accurate predictions enabled by AI can diminish the pressure exerted on healthcare professionals in overcrowded departments. The major requirements of AI-driven triage systems include strong data processing capability, real-time feedback mechanisms for clinicians, and Natural Language Processing (NLP) methods to ingest unstructured medical data \cite{DaCosta2025}. Everett et al. \cite{Everett2026} studied the integration of LLMs in real-world physicians' diagnostic workflow, evaluating two options: \textit{AI-as-first-opinion}, in which the clinician sees the LLM's diagnosis before producing an analysis, and \textit{AI-as-second-opinion}, in which the clinician provides the analysis before viewing the LLM output. In both pipelines, the analysis of the expert and the LLM are merged in a synthesized view, highlighting points of agreement and disagreement. Critical medical situations require monitoring systems capable of continuously tracking patients’ physiological data and issuing timely alerts to support prompt clinical intervention. In this context, human involvement is essential to lower the number of false alarms and identify suspicious patterns AI models may overlook. An emblematic example is sepsis diagnosis, a time-sensitive task in which the patient's vital signs have to be carefully monitored to ensure early detection \cite{Zhang2024-1}.

Given how clinicians' and AI's skills in healthcare complement each other, there is a need for systems that determine  whether a task should be assigned to the AI model or a clinician. Complementarity Driven Deferral to Clinical Workflow (CoDoC) \cite{Dvijotham2023} is an example of such a delegation system: using the confidence score from a predictive AI model, a deferral AI decides whether to rely on the model’s prediction of a medical image or instead hand the case over to a clinician. Evaluated on the tasks of detecting breast cancer and diagnosing tuberculosis, CoDoC improves specificity and reduces the number of required human reads compared with human-only single- or double-read workflows. Strong et al. \cite{Strong2025} designed a guided deferral system composed of an instruction-tuned LLM which decides whether to defer uncertain decisions to a human-in-the-loop, providing, in that case, a textual guide highlighting the main reasons for and against the diagnosis of a certain disorder.

\subsubsection{Limitations of HAIC in Healthcare} \hfill\\\\
\textit{Interpretability and Explainability.} In many applications, AI partners are neural-only models lacking interpretability, i.e., a transparent decision-making model, and explainability, i.e., the integration of techniques aiming to justify the model's prediction.  Clinicians interacting with black-box models may be unable to make accurate diagnoses and may fall into overreliance if they uncritically trust AI-generated results, or underreliance if they remain skeptical of the system's recommendations. Moreover, the lack of clear insight into the AI decision-making process may increase the workload of healthcare professionals, who must independently reconstruct the rationale underlying the system’s outputs. \\

\textit{Integration with existing workflows.} The widespread adoption of HAIC in healthcare is limited by the existence of human-only medical workflows that are resistant to changes and AI integration \cite{Olawade2026}. Fast-paced clinical environments already impose a substantial cognitive overload on clinicians, who are often reluctant to engage in time-consuming interactions with AI systems. To overcome this problem, more investigation is needed in designing novel human-AI frameworks and interfaces that support seamless collaboration. \\

\textit{Incompleteness of data-driven approaches.} Data-driven methods highly depend on the quality of data, which must be maintained sufficiently high for the entire duration of the HAIC system lifecycle. This process poses a substantial effort for dataset curators; it is error-prone and subject to change. For example, the patient population may change over time, and novel diseases may emerge and significantly impact data distribution. If the data are not representative of the target population, certain demographic groups or medical conditions may be underrepresented, leading models trained on such data to perform poorly in those cases. Annotation errors, ambiguities, and biases in the training data can also negatively affect the performance and reliability of decision-making models. \\

\textit{Cost of AI healthcare infrastructure.} AI systems for healthcare delivery often run on small devices equipped with limited hardware. To make AI solutions practically usable on resource-constrained platforms, novel lightweight algorithms are to be developed, with a particular focus on reducing training and inference cost. The sustainability of an AI algorithm should be assessed not only at deployment, but also in relation to the long-term maintenance and updates required to conform to shifts in data distributions and changes in the operating environment.

\subsection{HAIC in Cybersecurity} \label{sub:cyber}
In cybersecurity, HAIC refers to the collaboration between human analysts and AI-driven systems during cybersecurity operations. AI can continuously monitor vast amounts of data generated by cyber-assets, identify patterns leading to the discovery of threats, and automate incident response. On the other hand, experts use professional knowledge to evaluate attack scenarios and make well-informed decisions. The complementary strengths of AI and humans can enhance the overall accountability of defense operations spanning from intrusion detection to threat intelligence \cite{Sarker2024}.

Beyond harnessing the fruitful synergy of AI and humans, HAIC in cybersecurity must address challenges specific to the domain. A particularly important issue is the continuously evolving nature of cyber attacks and the rapid emergence of novel threats, which may quickly undermine the efficacy of conventional AI systems. In fact, common detection algorithms are trained on security logs that reflect only previously observed incidents and therefore lack information about new attack patterns. Human creativity is fundamental to recognizing suspicious data that may indicate the presence of intrusions or malware that are not included in the training data.

A frequently used approach to realize HAIC in cybersecurity relies on the integration of human feedback in the model workflow. In several studies, active learning is adopted to train detection models with a minimal set of data labeled by cyber analysts, reducing the overall labeling cost without sacrificing accuracy. Kim et al. \cite{Kim2024} proposed a human-in-the-loop intrusion detection system that employs an alert prioritization policy refined through human feedback using an AL scheme. In particular, humans produce an investigation outcome for each prioritized alert that is used to update policy beliefs; then sequential hypothesis testing is adopted to choose the hypothesis that best explains investigations and generated alerts. Evaluated against a static prioritization algorithm using a real-world attack graph, the developed method achieved a 79\% reduction in time to detection. Varandi et al. \cite{Varandi2024} experimented with several AL techniques for classifying network intrusion detection alerts and proposed a labeling pool initialization algorithm that achieves better performance than supervised models trained on the entire dataset. AL approaches are also well-suited for reducing labeling cost in scenarios involving data distribution shifts, as shown in \cite{Guerra-Manzanares2024}, where they have been applied to Android malware detection using data collected over a 7-year time frame, partitioned into timestamp-based chunks to support incremental model training.

HAIC can play a significant role in Security Operations Centers (SOC) by reducing human workload through the use of AI for repetitive tasks and preliminary investigations, while ensuring that cyber analysts receive sufficient and relevant information to support accurate decision-making. Building on this perspective, Tariq et al. \cite{Tariq2025-2} introduced the $\mathcal{A}^2\mathcal{C}$ framework, a conceptual model for HAIC in cybersecurity tasks within SOCs. The proposal emphasizes the importance of fostering situational awareness in both human and AI agents, enabling them to recognize their own strengths and limitations as well as those of their teammates, while also supporting adjustable levels of autonomy. The practical value of $\mathcal{A}^2\mathcal{C}$ for deployment in SOCs is further demonstrated through its application to an alert prioritization and validation use case. Beyond alert classification, another major challenge is to correlate them in order to identify when a security incident has occurred, and its impact extends to multiple data sources. Freitas et al. \cite{Freitas2024} designed GraphWeaver, a human-in-the-loop incident graph mining algorithm that can extract correlations between billions of security alerts. Threat researchers are closely engaged in the process, as they can continuously adjust the parameters and procedures used for correlation extraction and fine-tune the workflow to reduce the number of false correlations. The rapid growth of LLMs and AI agents introduces new possibilities for HAIC across a broad range of SOC functions \cite{Srinivas2025}. Large Language Models and agentic pipelines can augment cyber-analysts' work in many ways, for example, summarizing event logs into concise reports, or performing alert triage and threat intelligence. 

\subsubsection{Limitations of HAIC in Cybersecurity}\hfill\\\\
\textit{Human factors.} A major drawback of many HAIC cyber-systems is that they overlook human factors and their impact on performance. Stress levels and physiological conditions can substantially affect performance in cybersecurity tasks, yet these factors are rarely monitored to inform processes such as task allocation and alert management. The neglect of human factors contributes to unbalanced workloads in SOCs and to the resulting phenomenon of alert fatigue among security analysts. \\

\textit{Evolving threat landscape.} Classical AI data-driven approaches struggle to dynamically adapt to novel threats an organizational asset may be exposed to. Although AL techniques represent a first step towards adaptability, they require constant retraining of the models. This gap should be addressed through further research into continuous learning algorithms that are robust to distribution shifts and can operate with minimal human supervision. \\
 
\textit{Adjustable Autonomy.} Adjustable autonomy, i.e., the capability of a system to change its operation mode from full automation to collaboration, is underdeveloped in cybersecurity systems. The $\mathcal{A}^2\mathcal{C}$ framework represents a first step towards embedding this feature in real-world SOCs; however, further research is required. Effective HAIC frameworks for cybersecurity should support multiple working modes and dynamically adapt them according to human availability and AI capabilities.

\subsection{HAIC in Robotics}
The large diffusion of AI in physical systems, a phenomenon known as AI embodiment, has opened new possibilities in human-robot collaboration \cite{Semeraro2023}. Applications in which human operators are assisted by AI-enabled robots operating in the same workspace, referred to as \textit{cobotics}, have been shown to speed up productivity in manufacturing and reduce human cognitive and physical burden. Despite its effectiveness in industrial settings, close collaboration between humans and robots, particularly in the absence of physical barriers, raises safety concerns regarding the chance of collision and injuries for humans. The implementation of safety measures is imperative to minimize the probability and the severity of incidents in human-robot collaboration \cite{Li2024}.

In manufacturing, manipulator robots equipped with AI capabilities are used to assist humans in collaborative tasks. These include the assembly of complex objects by means of sequential steps, the exchange of items, the transport of objects, and the physical alteration of materials. In the majority of these cases, robots need to understand and anticipate human actions in order to offer assistance in the joint task. Situational awareness skills are essential to recognize the current step of the assembly or decide which object to hand over to the user. Zhang et al. \cite{Zhang2024-2} developed a skeleton-RGB integrated model for human action recognition that can distinguish highly similar actions from video streams and support online human action prediction. After acquiring human intention, the robot selects a dynamic response for assembling a worm-gear decelerator. Differently from other works, Kedia et al. \cite{Kedia2024} studied the interdependence between human intent and actions taken by the robot and built InteRACT, a transformer-based intent prediction model. To compensate for the scarce availability of human-robot interaction datasets, the authors conducted pretraining on human-human interaction data and a subsequent fine-tuning on human-robot interaction data.

Some HAIC methods adopted in robotics leverage human feedback to train robots in specific tasks. Following this approach, Deng et al. \cite{Deng2024} realized a collaborative scheme called \say{learning by doing}, in which visual knowledge captured by human operators wearing AR glasses is used to guide the learning of a disassembly robot through a Deep Active learning loop. The vision capabilities of the robot are iteratively refined by collecting data labeled by humans during the \say{doing} phase, and the operational framework was validated through a task involving the removal of screws from an EoL EV battery. Human expertise can also be transferred to robots through demonstrations of how the task of interest should be performed. Learning from demonstration (LfD) constitutes a viable approach when programming and end-to-end learning methods are not suitable for the given application. Liu et al. \cite{Liu2025} used a single third-person demonstration of a human performing an assembly task recorded with an RGB-D camera to teach the robot. The pipeline includes the extraction of relevant features from the captured video and an algorithm to imitate human movements and generalize them to novel situations.

A mature HAIC system for robotics should incorporate an adaptive task-allocation framework that allows humans and robots to exchange roles during task execution and flexibly perform different operations without being constrained by a rigidly predefined workflow. Noormohammadi-Asl et al. \cite{Noormohammadi-Asl2025} introduced a task allocation and scheduling framework for collaborative execution that takes into account human preferences for leading or following and changes in performance, focusing on improving team outcome and task completion time in kitting and pick-and-place scenarios. Specifically, the human and the robot have the agency to allocate tasks to themselves and to the other agent asynchronously; the robot performs task allocation and scheduling after estimating the human's changes in preference and performance, and can assume the leading role when the human's performance is poor. Ng et al. \cite{Ng2024} applied a diffusion model to learn a co-policy, wherein the robot predicts the future action of both the human and itself and executes the latter. Co-policy learning used demonstrations from human-human interaction datasets and led to significant improvements in collaborative table-carrying tasks with respect to state-of-the-art imitation learning methods.

%Human-intent prediction and robot motion planning are complementary skills in a HAIC system. In \cite{Kedia2023b}, authors proposed ManiCast, a novel framework integrating a cost-aware human forecast model and a Model Predictive Control (MCP) planner. The forecasts are directly used to model the robot's planner cost function in collaborative tasks such as reactive stirring, object handovers, and collaborative table setting with a 7-DoF robot arm. Following the same paradigm, \cite{Abilkassov2025} et al. combined deep learning-based human hand trajectory forecasting and heuristic optimization for robotic task sequencing to avoid collisions in a collaborative scenario. 

Classical approaches adopted in human-robot collaboration are often limited to specific tasks and struggle to adapt to novel environments. Foundation models, such as LLMs, retain a high potential to generalize reasoning capabilities and improve the flexibility of HAIC systems in robotics \cite{Kim2024-2}. Ji et al. \cite{Ji2024} designed a collaborative framework composed of a perception module, powered by a speech recognition model and a Vision Foundation Model (VFM), an LLM-based reasoning module, and an execution module. In this workflow, human language instructions are transcribed into text and combined with a description of the surrounding environment to form a prompt for an LLM. The LLM then produces low-level code that can either be executed directly on the robot or further improved through several rounds of interaction with a human operator. Ranasinghe et al. \cite{Ranasinghe2025} proposed CogniVera, a dual-agent system for collaborative task assembly execution that enables robots to communicate vocally with humans. To mitigate system-generated errors and hallucinations, CogniVera includes a Cognitive Processing Unit hosting a Generation Agent that produces function calls for robot execution and other responses from human input, and a Validation Agent to verify that generated output satisfies the user request. %Experimental validation in the collaborative assembly yielded an error correction rate of nearly 98\% and high task completion compared with a baseline without the Validation agent. 

\subsubsection{Limitations of HAIC in Robotics} \hfill\\\\
\textit{Limited Shared Autonomy.} Although flexible task allocation, leadership switches, and mutual adaptation have been explored in collaborative robotics, collaboration is still limited by naive assumptions such as task-specific constraints and predefined roles. Moreover, human–robot interaction is exposed to risks arising from frequent robotic errors, which may range from sensing failures to inappropriate interruptions of human activities \cite{Cao2025}. Enabling robots to autonomously detect and correct their own errors remains an open challenge. \\

\textit{Real-time human intent prediction.} A second limitation lies in the difficulty of accurately predicting human intentions in real time. Many collaborative robots rely on machine-learning models to infer what the human partner is going to do next, such as in assembly, handover, or shared manipulation tasks. However, human behavior is variable, context-dependent, and sometimes ambiguous: different intentions can produce similar motions, while the same task can be performed in different ways by different users \cite{Lidard2024}. \\

\textit{Poor generalization skills.} Another important challenge is the limited robustness and generalizability of AI-based perception in human–robot collaboration. Vision-based systems are widely used to recognize gestures, actions, objects, and task states, but they may be sensitive to occlusions, lighting changes, camera viewpoints, and sensor noise. This aspect severely limits the transferability of models trained in controlled laboratory conditions to real industrial environments, where the robot must cope with cluttered scenes, unexpected human behavior, and changing task configurations \cite{Robinson2023}. \\

\textit{Safety concerns.} Errors in motion planning for a robot could result in severe harm for humans. However, cobots are usually tested in simplified conditions, and their robustness under noisy sensing, unpredictable human movements, and unusual environments remains unverified. Before a deployment in real-world scenarios, HAIC applications in robotics should be tested with explicit consideration of the potential consequences of failures in perception and decision-making.

\subsection{Why Is Neurosymbolic AI Relevant to HAIC?}
There are several reasons for adopting NeSy AI in collaborative contexts, and these motivations span all the identified application domains.

One of the major potential contributions of NeSy AI lies in enhancing the interpretability of the decision-making process. The use of a symbolic interface made of human-understandable symbols establishes a \textit{lingua franca} that naturally facilitates interaction between humans and AI systems \cite{Kambhampati2022}. In the medical domain, it is fundamental to accurately support diagnosis and decisions with evidence and findings from patients' observations and clinical history. The integration of a symbolic component, such as a knowledge graph defining causal relationships between symptoms and diseases, can add an interpretability layer to prediction models used in healthcare \cite{Wu2025}. Similarly, cybersecurity tasks such as anomaly detection, alert prioritization, and incident response can be handled with \say{intelligible} models that integrate explicit representations of cyber-knowledge. Cybersecurity Knowledge Graphs (CKG) \cite{Sikos2023} are commonly used to transform available knowledge into an ontology-based form that can capture different aspects of a cyber-system, such as the network infrastructure, the flow of information, attack paths, or a vocabulary of cybersecurity concepts. The CKG can be iteratively refined by experts when new knowledge becomes available, ensuring that interpretations are not obsolete. Enhanced interpretability can positively influence users’ trust in AI systems. The opportunity to inspect the reasoning underlying model outputs enables users to make more informed decisions about whether to accept or reject AI recommendations, thereby mitigating both overreliance and underreliance. Moreover, in many HAIC applications, interpretability is not regarded as an optional add-on but as an essential requirement for ensuring both system and human safety, which may be addressed through a neurosymbolic approach.

Neurosymbolic integration can also compensate for the absence of high-quality and representative data. The incorporation of curated medical ontologies, knowledge graphs and clinical constraints is useful to describe patterns that can be learned only partially from data. For example, neural models trained on datasets where patient groups or medical conditions are unevenly represented are likely to fail at learning robust representations for minority classes. Therefore, incorporating rules that help identify such cases can be beneficial. The same condition emerges in cybersecurity systems that are mostly able to detect only known attacks seen during training: the inclusion of symbolic mechanisms to detect novel attacks or variations of existing ones can make intrusion detection systems more robust. The limits of pure data-driven methods are also evident in robotics. Many current systems are strongly tied to specific tasks or environments because they learn direct mappings from perception to action. A neurosymbolic architecture could instead separate low-level perception from high-level task reasoning. Neural models may recognize human actions, objects, gestures, speech commands, or motion trajectories, while symbolic representations could encode the task hierarchy, object affordances, preconditions, temporal dependencies, and goal states. In this way, the robot would not merely classify the human’s actions, but also reason about their meaning within the context of the collaborative task. Symbolic reasoning can also enhance transferability to new scenarios, as part of the acquired knowledge is encoded in reusable symbolic structures rather than being entirely embedded in task-specific neural parameters.

In evolving domains where data undergo frequent distribution shifts, neural-only models must be constantly retrained to deliver accurate performance. This fact imposes a high computational and maintenance cost on HAIC systems developed in cybersecurity and healthcare, which must adapt to novel threats and patient diversification, respectively. The retraining effort can be lightened by including a symbolic component that encodes part of the knowledge. Small updates in the knowledge bases can help to minimize the frequency of retraining and the amount of data needed, mitigating the problem of collecting up-to-date, high-quality data. Reducing the computational complexity of HAIC systems using Neurosymbolic AI is crucial both for meeting the real-time requirements of many applications and enabling deployment on computational devices with limited resources.

The realization of Adjustable Autonomy in HAIC scenarios can also benefit from Neurosymbolic AI. In cybersecurity and robotics, delegation decisions are often triggered by conditions related to performance or safety. These conditions can naturally be expressed as symbolic rules, thereby motivating the adoption of a neurosymbolic approach. More generally, NeSy AI may define safety guardrails blocking unauthorized and hazardous actions. In robotics, this integration could enhance system robustness by filtering neural uncertainty through explicit constraints, thereby reducing the likelihood that perception or prediction errors result in unsafe robot behavior. In healthcare, it could support compliance with clinical standards and established best practices.

\subsection{Neurosymbolic Human-AI systems}
Although NeSy AI has the potential to fill existing research gaps in HAIC, its current adoption is quite limited in this context. We review several works that present a neurosymbolic methodology within a hybrid collaborative application scenario.

Kalutharage et al. \cite{Kalutharage2025} presented a neurosymbolic model for anomaly detection in an IoT network that leverages neural network-based anomaly detection, the SHAP method, and a knowledge graph. Data instances are passed through an autoencoder, recording the reconstruction loss and the SHAP values. SHAP identifies the top influential features, and for those features, the model checks whether the corresponding original feature value exceeds the threshold stored in the knowledge graph, reporting an anomaly in that case. The knowledge graph is created from normal and known attack data, and it is subsequently augmented with expert knowledge. Notably, the framework also provides a direct mapping between anomalous features and violations of Confidentiality, Integrity and Availability (CIA), and an LLM-generated explanation based on relevant insights from MITREATT\&CK framework.

Mandlekar et al. \cite{Mandlekar2023} introduced Human-In-The-Loop Task and Motion Planning (HITL-TAMP), a robotics data-collection and learning system for collaboration between a robot and a human using teleoperation. A TAMP-gated policy, based on a logic-based planning framework PDDLStream, prompts the human to execute only task segments involving difficult contact-rich operations. Collected demonstrations are employed to train a policy through imitation learning for human-controlled segments. Unlike common teleoperation systems, the deployed framework does not require humans to be present throughout the entire task execution, but involves them only at critical bottlenecks, allowing a single person to supervise and command a fleet of robots in parallel.

Singh et al. \cite{Singh2025} combined an LLM with a knowledge graph refined through human input to decompose generic cooking and cleaning tasks into a sequence of executable actions. Specifically, the LLM is prompted with task-specific information and an example of execution. The generated program is evaluated by verifying whether its ontology entities correspond to those in the knowledge graph, and any mismatches are corrected through replacements. Programs that contain errors after this phase are revised by re-prompting the LLM and then interactively refined through human input. LLM+KG+Human achieved better task success rates and less token consumption than the LLM deployed in isolation.

Shyalika et al. \cite{Shyalika2026} developed CausalTrace, a neurosymbolic causal-analysis module inside an industrial manufacturing copilot called SmartPilot. The tool supports neural-based causal discovery from process data, root cause analysis, and counterfactual reasoning, while grounding these outputs in symbolic domain knowledge encoded through manufacturing ontologies, knowledge graphs, and causal graphs. In this setting, the human operator remains in the loop by querying the system, inspecting proposed causal chains, validating or refining causal hypotheses, and using the generated explanations to decide which process variables or components should be adjusted.

Parikh et al. \cite{Parikh2026} evaluated a human-AI teaming workflow for oncology clinical-trial prescreening, where clinical research coordinators used a neurosymbolic AI system to assess patient eligibility from retrospective electronic health records. The neural component consists of clinical language models that process longitudinal, unstructured EHR text and extract eligibility-relevant information, such as biomarkers, cancer stage, treatment history, response status, and exclusionary conditions. These extracted facts are then organized according to the symbolic structure of trial eligibility criteria, which encode inclusion and exclusion rules at the criterion level. In the collaborative workflow, the AI does not replace the coordinator but pre-abstracts and structures evidence from the chart, while the human reviewer inspects this evidence, resolves ambiguities, and makes the final prescreening decision.

%% file: img/hexagons.tex
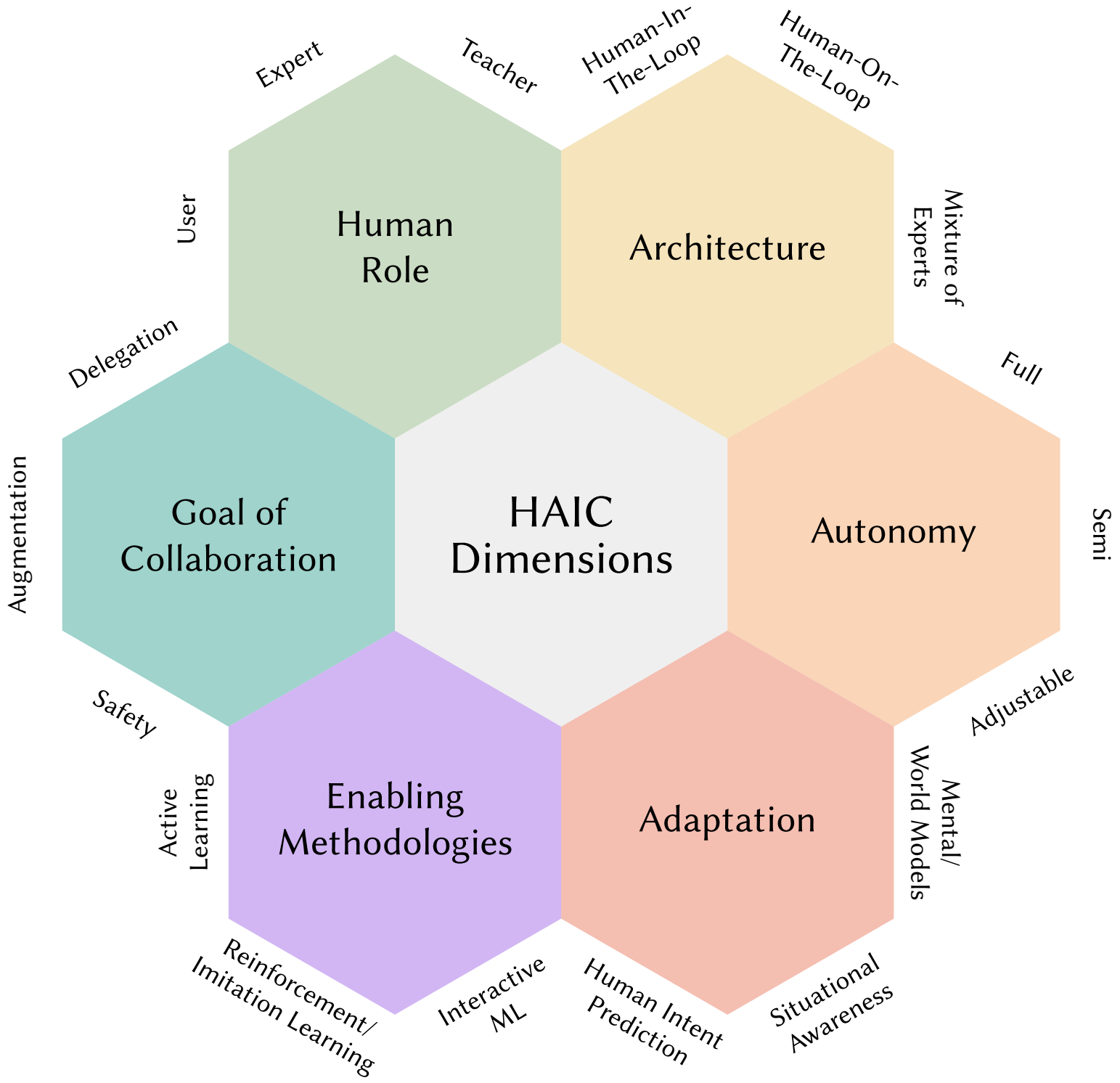
\begin{figure}[ht!]
\centering
\begin{tikzpicture}

% Distance between the centers of two edge-sharing hexagons
\pgfmathsetmacro{\HexDistance}{sqrt(3)*\HexRadius}

% Central hexagon
\coordinate (C) at (0,0);
\Hexagon{C}{CentralColor}{ \Large HAIC \\ \Large Dimensions }

% Coordinates of the six surrounding hexagons
\coordinate (H1) at ($(C)+(180:\HexDistance)$);
\coordinate (H2) at ($(C)+(120:\HexDistance)$);
\coordinate (H3) at ($(C)+(60:\HexDistance)$);
\coordinate (H4) at ($(C)-(180:\HexDistance)$);
\coordinate (H5) at ($(C)-(120:\HexDistance)$);
\coordinate (H6) at ($(C)-(60:\HexDistance)$);

% Surrounding hexagons
\Hexagon{H1}{HexagonOne}{Goal of \\ Collaboration}
\Hexagon{H2}{HexagonTwo}{Human \\ Role}
\Hexagon{H3}{HexagonThree}{Architecture}
\Hexagon{H4}{HexagonFour}{Autonomy}
\Hexagon{H5}{HexagonFive}{Adaptation}
\Hexagon{H6}{HexagonSix}{Enabling \\ Methodologies}

% Labels along the external edges
\HexEdgeLabel{H1}{2}{Delegation}
\HexEdgeLabel{H1}{1}{Augmentation}
\HexEdgeLabel{H1}{6}{Safety}

\HexEdgeLabel{H2}{1}{\quad \quad User}
\HexEdgeLabel{H2}{2}{Expert}
\HexEdgeLabel{H2}{3}{Teacher}

\HexEdgeLabel{H3}{2}{\quad \quad Human-In-\\\quad \quad The-Loop}
\HexEdgeLabel{H3}{3}{Human-On-\\The-Loop}
\HexEdgeLabel{H3}{4}{Mixture of \\Experts}

\HexEdgeLabel{H4}{3}{\quad \quad Full}
\HexEdgeLabel{H4}{4}{Semi}
\HexEdgeLabel{H4}{5}{\quad \quad Adjustable}

\HexEdgeLabel{H5}{4}{Mental/ \\ World Models}
\HexEdgeLabel{H5}{5}{Situational \\ Awareness}
\HexEdgeLabel{H5}{6}{\quad \quad Human Intent \\ \quad \quad Prediction}

\HexEdgeLabel{H6}{1}{Active \\ Learning}
\HexEdgeLabel{H6}{6}{Reinforcement/ \\ Imitation Learning}
\HexEdgeLabel{H6}{5}{Interactive \\ ML}

\end{tikzpicture}
\caption{HAIC dimensions}
\label{fig:haic_dim}
\end{figure}

%% file: section_framework.tex
\section{NeSy-HAIC framework} \label{sec:framework}
\subsection{Overview}
\begin{figure}[htpb]
    \centering
    \includegraphics[width=1\linewidth]{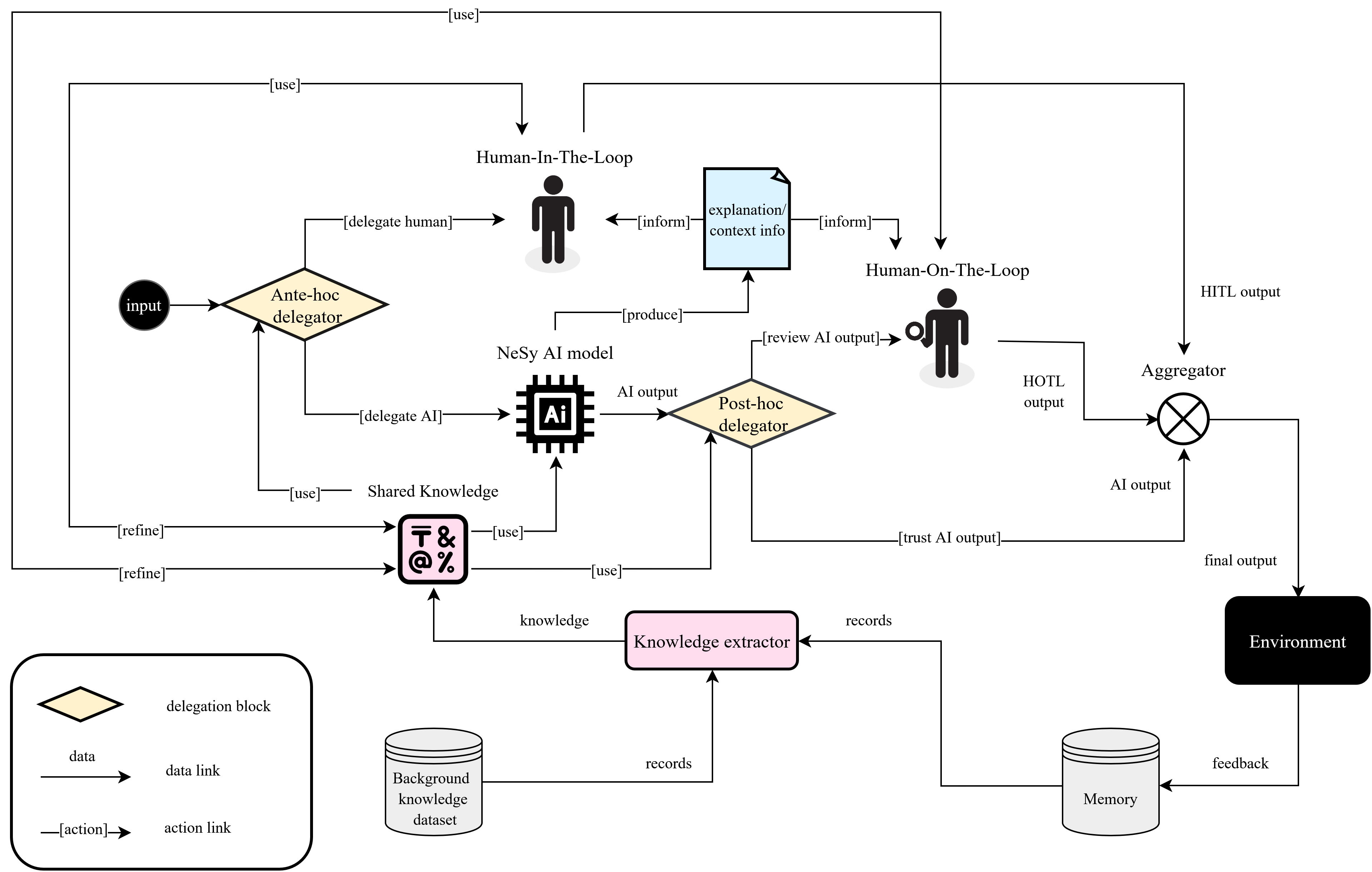}
    \caption{NeSy-HAIC framework}
    \label{fig:schema_all}
\end{figure}

In this section, we present an overview of our neurosymbolic framework for HAIC, called NeSy-HAIC. As far as we know, this is the first attempt to explicitly model HAIC in combination with Neurosymbolic AI. Figure \ref{fig:schema_all} shows a visual representation of the proposed framework, which serves as a high-level view of how the inference process is performed. It integrates multiple AI-based components and two human actors with different roles, allowing human and AI parties to engage in varying types of interactions. In fact, our architecture is designed to be as general and expressive as possible, since our main goal is to characterize the majority of HAIC scenarios without delving into implementation details and domain-specific nuances. The resulting modularity offers an advantage for modeling a wide range of collaborative tasks and makes it possible even to remove certain components when they are not necessary for specific applications.

We outline the supported components and decision-making modes in our framework, along with enabling methodologies for the actual framework implementation. 

\subsubsection{AI-based components}\hfill\\
\begin{itemize}
    \item Ante hoc delegator. This AI-based delegation module establishes whether the task should be assigned to the human expert or the NeSy AI model. We call it \textit{Ante hoc} because the delegation decision precedes the execution of the task by either the human or the AI.
    \item Post hoc delegator. It determines whether the output generated by the NeSy AI model is acceptable or human intervention is required. In the latter case, the Human-On-The-Loop (HOTL) is invoked. This module is named \textit{Post hoc} because the decision to delegate occurs only after task execution by the AI.
    \item NeSy AI model. It leverages symbolic knowledge and neurosymbolic integration to execute the task of interest. Compared with a purely neural network–based approach, it enhances interpretability by combining data-driven learning with symbolic reasoning and can also produce human-readable explanations that clarify the rationale behind decisions.
    \item Knowledge Extractor. This component derives new symbolic knowledge from past interactions with the environment, collected in a memory, or from a Background Knowledge dataset. The goal of this module is to perform Automatic Knowledge Discovery and expand the system’s symbolic knowledge base without direct human involvement.
\end{itemize}

\subsubsection{Decision-making modes} \hfill\\
Our framework encompasses two different human actors corresponding to two alternative decision-making modes, namely Human-In-The-Loop (HITL) and Human-On-The-Loop (HOTL). These modes, along with their possible combinations, are well-suited to handle diverse HAIC tasks and various forms of human-AI interaction.

\begin{itemize}
    \item Human-In-The-Loop. In the HITL mode, a human expert can be delegated to execute the task in place of AI. Specifically, the Ante hoc Delegator establishes whether delegating to the expert is more convenient than delegating to the NeSy AI model and subsequently assigns the task to the selected entity. To maximize expected performance, the Ante hoc delegator defers to the performer who is most likely to yield the best results. In some cases, delegation is not only expected to improve outcomes but also to minimize the costs associated with human intervention, for example, by limiting how often the human can be deferred.
    \item Human-On-The-Loop. In the HOTL mode, a human expert intervenes to review and eventually modify the output of the AI model, which may produce errors. This action is intended to compensate for individual shortcomings of the NeSy AI model, which may arise, for instance, in the presence of data distribution shifts or when the symbolic knowledge is incomplete. The delegation decision is taken by the Post hoc delegator based on the output of the NeSy AI model. A typical way to determine when to delegate is to evaluate the AI output’s confidence and trigger the human whenever the confidence score is low.
\end{itemize}

These decision-making modes are not mutually exclusive and can be combined to support hybrid HIT–HOTL solutions. For example, the human experts in HITL and HOTL modes can either rely on a common knowledge base or exhibit complementary skill sets, increasing the overall robustness of the system. Given the output of the HITL mode expert, the output of the NeSy AI model, and the HOTL mode expert output, an aggregation or selection strategy is adopted to produce the final output. By doing so, an HAIC system built on our framework can simultaneously leverage decisions made by the human alone, the AI alone, and the AI followed by human revision.

\subsubsection{Symbolic Knowledge and Neurosymbolic Integration} \hfill\\
To support effective neurosymbolic integration, our framework incorporates a dedicated symbolic knowledge management system. The Shared Knowledge block consists of a collection of symbolic artifacts expressed in a formal language, e.g., knowledge graphs, ontologies, and logical rules, that is accessible to every human and AI-based module within the framework. From this perspective, symbolic knowledge acts as the communication interface between humans and AI components.

For the construction and maintenance of the Shared Knowledge, we define the following operations:
\begin{itemize}
    \item Initialization. Symbolic knowledge is extracted by applying data-driven methods to a Background Knowledge dataset or provided by a human.
    \item Human refinement. At any point during the system lifecycle, the humans in HITL and HOTL mode may update the Shared Knowledge leveraging their expertise.
    \item Automatic Knowledge Discovery. Based on environmental feedback or other additional data, the Knowledge Extractor can automatically discover new symbolic knowledge and integrate it into the Shared Knowledge.
\end{itemize}

Once available, Shared Knowledge serves multiple functions within our framework. The NeSy AI model can leverage it to perform the task of interest, while the Ante hoc and Post hoc Delegators can exploit it to ensure that delegation strategies comply with domain-specific rules. 

\subsubsection{Closed-loop and Open-loop configurations} \hfill\\
To model the presence of a feedback loop within our framework, we introduce an environment module that returns feedback to the HAIC system. This configuration is typical of closed-loop tasks, in which the effects of the system's outputs are observed and subsequently influence future decision-making, e.g., a robot using a LIDAR sensor to perceive its surroundings and navigate an indoor environment. In our framework, environmental feedback is stored in a memory and subsequently processed by the Knowledge Extractor to automatically construct new symbolic knowledge.

By removing the Environment and the Memory blocks, our architecture easily adapts to open-loop tasks in which the feedback loop is absent, e.g., an intrusion detection system that classifies incoming network traffic as either benign or malicious.

\subsection{Enabling methodologies}

\subsubsection{Ante hoc and Post hoc delegator} \hfill\\
Learning to Abstain (L2A) \cite{Punzi2026} is a common AI technique used to refrain from making a prediction when it is likely to be wrong. L2A systems are further classified into Learning to Reject (L2R) systems, which learn to refuse AI output under uncertainty conditions, and Learning to Defer (L2D) systems, which are trained to delegate to a downstream human expert. Despite this distinction, L2R can be regarded as a special case of L2D, since both paradigms seek to learn a predictive model together with a rejection function that determines whether the model’s prediction should be accepted or deferred to a human decision-maker. Two principal types of rejection mechanisms can be distinguished: novelty rejection, which discards samples that significantly deviate from the training data distribution, and ambiguity rejection, which rejects samples that lie near the decision boundary of the predictive model. Novelty rejection methods are typically model-agnostic and can be implemented using statistical techniques developed for Outlier Detection or Anomaly Detection \cite{Coenen2020}. In our setting, the Ante hoc delegator can be implemented as a novelty detector that bases its rejection decision solely on the input. In contrast, ambiguity rejection methods are tightly coupled with the predictive model, making them particularly suitable for the Post hoc delegator. For example, stage rejectors use the predictor’s output to compute a confidence score subsequently compared against a predefined threshold, and joint rejectors train the predictor and the rejector together, treating the rejection option as an additional class.

Instead of replacing or deferring humans, Learning to Complement (L2C) proposes an alternative delegation model using AI models that specialize in decision regions where human performance drops. This complementary approach makes delegation robust to individual human failures, which are typically overlooked in L2D methods. Although only a limited number of studies have investigated this paradigm \cite{Wilder2020, Zhang2024, Hemmer2022}, L2C provides an opportune paradigm for implementing both the Ante hoc and Post hoc delegator.

Alternatively, delegation can be formulated as a reinforcement learning problem, as proposed by Fuchs et al. in \cite{Fuchs2023}. In their architecture, a delegation manager receives environmental observations and delegates control either to the human or to the autonomous AI system based on the expected cumulative reward. Using a different approach, the same authors designed an RL-based delegation manager which intervenes only when domain-specific constraints are violated \cite{Fuchs2024}.

The previous techniques for realizing delegation are mostly neural-only and do not leverage symbolic knowledge or provide explanations about the deferral process. We argue that NeSy AI can strongly contribute to the design of the Ante hoc and the Post hoc delegator, although neurosymbolic integration remains largely underexplored in this area. For instance, Pugnana et al. introduced Deferring Concept Bottleneck Models (DCBMs) \cite{Pugnana2026}, extending traditional CBMs with a deferral option for both the intermediate concepts and the task predictions. In this deferral system, the availability of interpretable concepts on which humans can intervene increases the overall explainability and trustworthiness.

\subsubsection{Neurosymbolic AI model} \hfill\\
Given the methodological diversity of NeSy AI, our framework does not prescribe a single neurosymbolic integration strategy but accommodates multiple solutions.  The choice of the NeSy AI model is highly influenced by the type of task, the format of symbolic knowledge, and other design dimensions for neurosymbolic systems described in Section \ref{sec:nesy}. For example, neurosymbolic reinforcement learning methods are particularly well suited to sequential decision-making tasks in a closed-loop configuration. When knowledge is represented as logical statements, the model may incorporate a symbolic solver; when it is encoded as a KG, graph-traversal algorithms and multi-hop reasoning may instead be employed. Depending on the application requirements, the selected architecture may either maintain a clear separation between the neural and symbolic components or integrate them tightly to enable end-to-end training.

\subsubsection{Knowledge Extractor} \hfill\\
The manual construction of symbolic knowledge by domain experts is a time-consuming and non-scalable process. Therefore, methods are needed to automatically extract symbolic knowledge from data and interactions with the environment. The enabling methodologies for the Knowledge Extractor vary depending on the symbolic formalism used to represent background knowledge and the symbolic formalism of the target output. Inductive Logic Programming (ILP) can induce hypotheses from examples and background knowledge expressed in logic, making it particularly appropriate when the target representation consists of interpretable rules. Modern neurosymbolic extensions of ILP, such as Differentiable ILP (dILP) \cite{Evans2018}, enable logical rules to be learned from noisy perceptual input by coupling classical ILP with neural modules. A second family of knowledge extraction methods includes techniques for the construction of ontologies or knowledge graphs from unstructured data sources. A notable example from this category is REBEL \cite{Cabot2021}, which directly produces subject–relation–object triples from raw text leveraging end-to-end sequence generation.

\subsection{Case studies}
To demonstrate the practical value of our framework, we present two concrete application examples in semi-autonomous driving and alert prioritization. We chose these tasks to ground our model in both HITL and HOTL settings, thereby demonstrating its versatility and suitability for diverse HAIC domains. 

\subsubsection{Semi-autonomous Driving} \hfill\\
Semi-autonomous driving harnesses human-machine interaction for controlling a vehicle and introduces human participation to compensate for deficiencies of fully autonomous driving systems, such as the expensive labeling requirements and the inability to cope with unknown situations \cite{Ning2022}. Hybrid driving frameworks can support different modalities of interaction. For instance, Blended Shared Control lets the human driver \say{co-drive} with the autonomous system by continuously merging the human and the automatic control input and opportunely fixing conflictual actions. On the other hand, in Haptic Shared Control, the human retains full control of the vehicle; however, the machine engages in supportive actions (e.g., steering for lane-keeping) that guide the driver. Other control mechanisms focus on cooperation and task distribution between the human and the machine, e.g., alternating between them in control of the vehicle. 

\begin{figure}[ht!]
    \centering
    \includegraphics[width=1\linewidth]{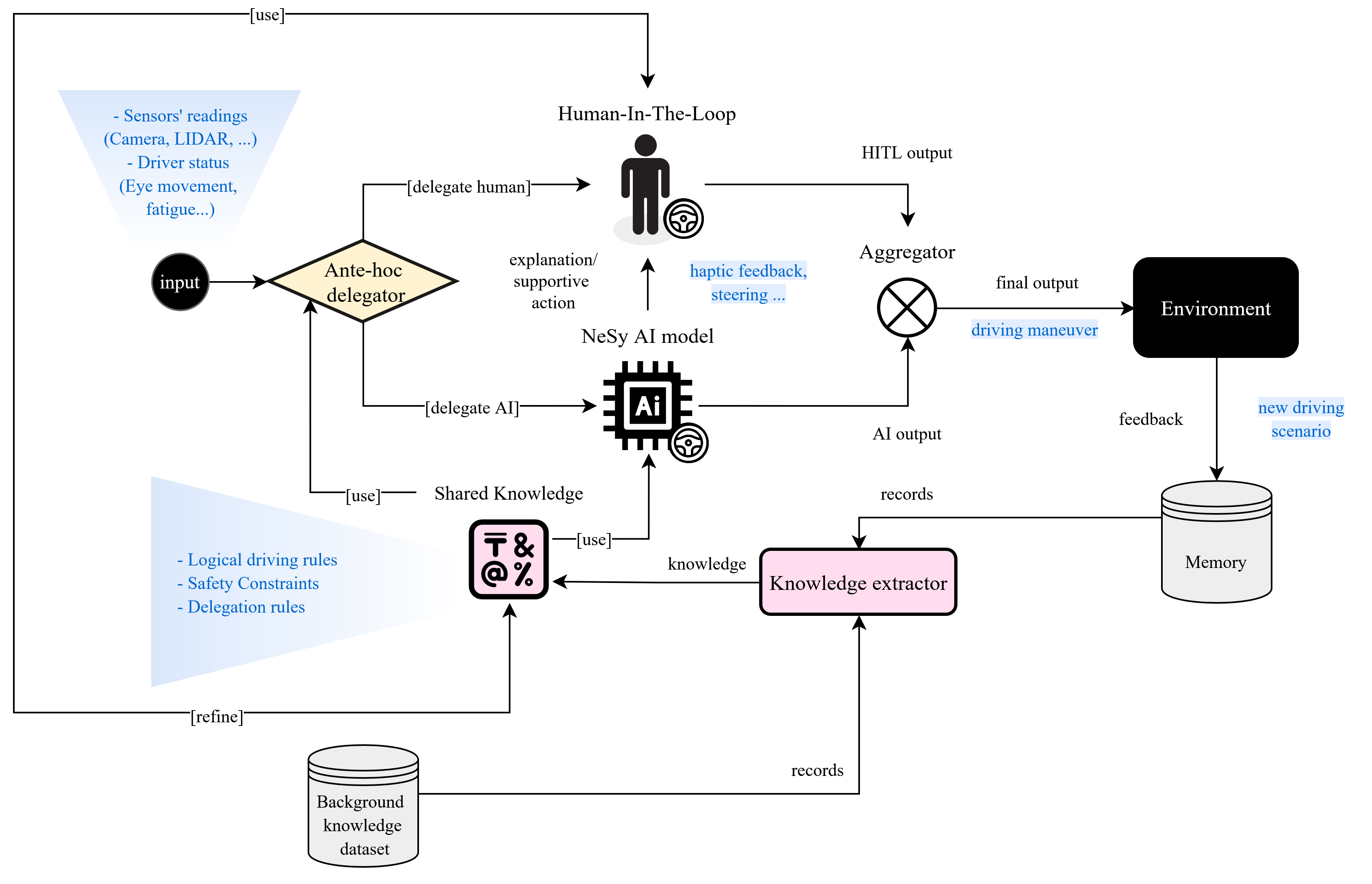}
    \caption{NeSy-HAIC framework using the HITL decision mode for a semi-autonomous driving task}
    \label{fig:hitl}
\end{figure}

Each of the previously discussed HAIC approaches can be regarded as a variation of the HITL decision mode and is therefore explicitly represented within our framework. Figure \ref{fig:hitl} shows a simplified view of our general framework, where only the components relevant to the HITL scenario are retained. These components are adapted to the semi-autonomous driving task.

The inputs to the system include information about the vehicle and its surroundings collected through sensors like cameras, LIDARs, gyroscopes, and Inertial Measurement Units (IMU). Advanced systems may also incorporate sensing capabilities to detect drivers' status, such as appearance, eye movement, and fatigue. Indeed, although they are often overlooked, human factors should also be considered in the driving policy. 

The Shared Knowledge collects symbolically represented driving rules and safety constraints as well as information supporting effective delegation.

\begin{align}
    Turn(X,Y) &:- \, Direction(X) \, \land \, Road(Y)\land \, Free(X,Y) \label{eq:sk1}\\
    AIControl &:- \, Fatigue(X) \, \land \, GreaterThan(X,limit) \label{eq:sk2}
\end{align}

Rules (\ref{eq:sk1}) and (\ref{eq:sk2}) provide a practical example of how this knowledge can be represented in first-order logic. The first rule specifies that the vehicle can execute a turn only if no approaching vehicles or obstacles are present in the intended turning direction, whereas the second requires control to be transferred to the AI system when the driver’s fatigue level exceeds a predefined threshold. Input information and Shared Knowledge are forwarded to the Ante hoc delegator, which determines whether control should be assigned to the human driver, the autonomous system, or both jointly. Specifically, the autonomous system is implemented as a neurosymbolic model that produces a control action as its output. As in Haptic Shared Control, this module may provide support to the driver through action suggestions or explanations. Finally, the selected action is executed in the driving environment.

Our framework illustrates how neurosymbolic integration offers multiple advantages in the semi-autonomous driving context. First, our system incorporates symbolic guardrails that constrain the autonomous driving agent's policy, avoiding or mitigating unsafe actions. This ensures that the NeSy AI model complies with explicit driving rules rather than relying exclusively on a performance objective that may give rise to unintended behavior. Moreover, a neurosymbolic driving policy may generalize more effectively to novel scenarios by leveraging driving rules that remain valid across different operating conditions. The second contribution of neurosymbolic integration lies in enhancing delegation through control mechanisms designed to ensure smooth and safe transitions between the human driver and the AI system. As encoded in Rule (\ref{eq:sk2}), control should be transferred to the AI system when the human driver is fatigued; conversely, the human driver should assume control when the AI vision system is faulty. These rules can be translated into a symbolic form and integrated into a neurosymbolic implementation of the Ante hoc delegator.

To illustrate our HITL framework in the context of semi-autonomous driving, we compare it with the hybrid driving system proposed by Fuchs et al. \cite{Fuchs2024}. The authors introduced a novel architecture based on an RL manager that delegates either the human or the AI agent to navigate traffic intersections. The manager is trained to jointly maximize the expected task performance and minimize the frequency of interventions, which are triggered whenever a manager's constraint is violated, e.g., a collision occurs. The desired behavior therefore encompasses not only reducing constraint violations but also maintaining the selected agent in control for extended periods. This system can be regarded as a specific instantiation of our architecture; in fact, the RL Delegation manager corresponds to the Ante hoc delegator. By framing the task of learning the manager's delegation policy as an Intervening Markov Decision Process (IMDP), the manager's interventions are confined to states violating predefined symbolic constraints, e.g., the vehicle exceeds a speed limit.

This approach provides a preliminary yet concrete example of how symbolic knowledge can guide delegation in human–AI driving scenarios by restricting interventions to specific states of a predefined IMDP. By explicitly defining the conditions under which the manager intervenes, the framework reduces the policy’s exploration space and, consequently, simplifies the delegation problem. However, although the exploration space is constrained, the manager remains a purely RL-based agent whose behavior is guided exclusively by the reward signal rather than also being governed by symbolic constraints. Similarly, the autonomous driving system does not exploit explicit symbolic driving rules and is implemented as a conventional RL agent. In contrast, our framework can also accommodate neurosymbolic implementations encoding such rules.

\subsubsection{Intrusion Detection and Alert Prioritization} 
\begin{figure}[ht!]
    \centering
    \includegraphics[width=1\linewidth]{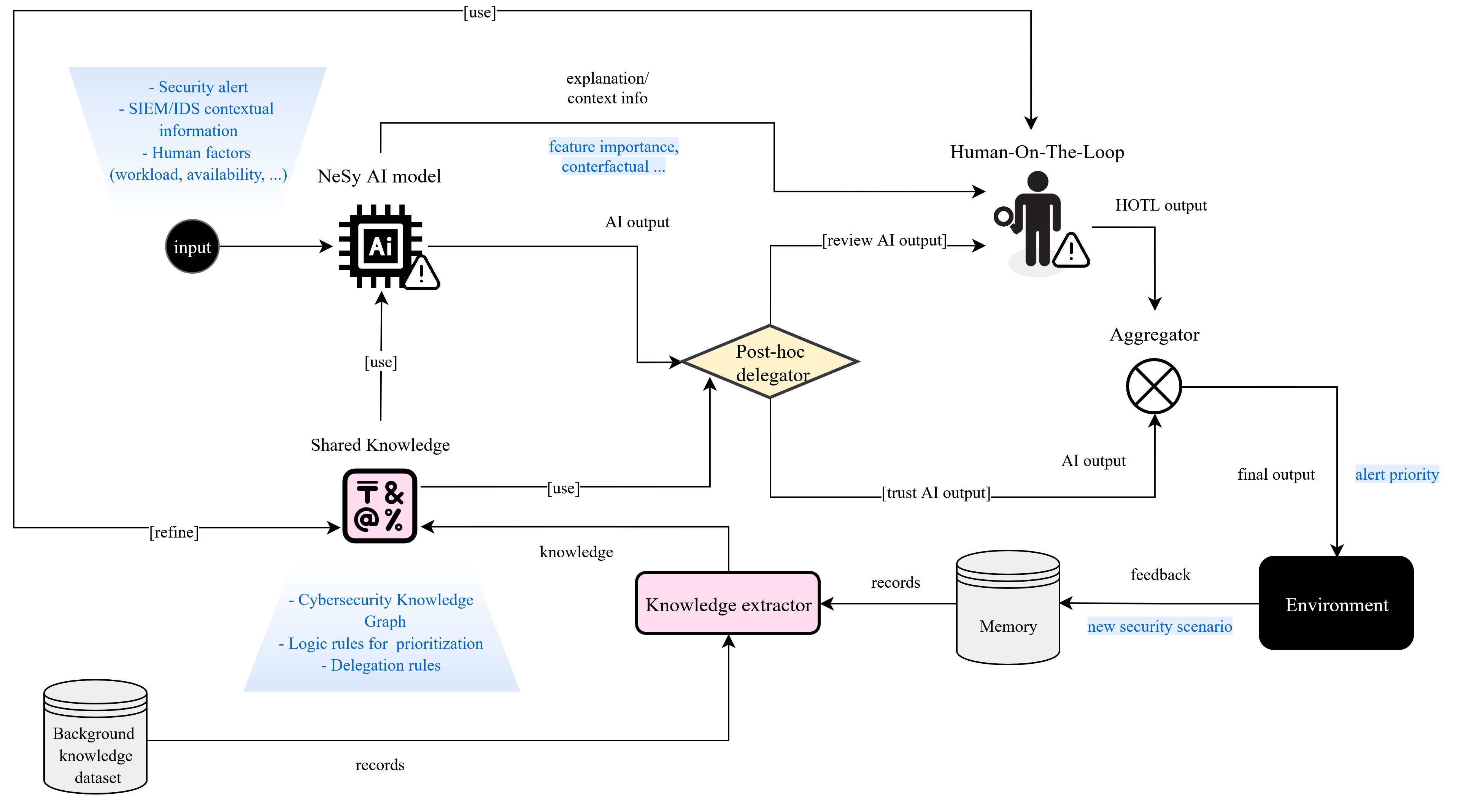}
    \caption{NeSy-HAIC framework using the HOTL decision mode for an alert prioritization task}
    \label{fig:hotl}
\end{figure}

\hfill\\
As described in Section \ref{sub:cyber}, cyber-defense represents a natural domain for the development of HAIC solutions in which experts team up with automated tools. In Security Operations Centers (SOCs), analysts are increasingly supported by AI-enabled technologies, including Intrusion Detection Systems (IDS) and Security Information and Event Management (SIEM) systems, to identify critical threats among high volumes of alerts \cite{Jalalvand2024}. The typical configuration of HAIC systems for alert prioritization involves an AI model that processes alert data, recognizes underlying patterns, and detects potential attacks, and a human analyst who conducts an in-depth investigation of the alerts prioritized by the AI system. This workflow enables analysts to assume a supervisory role while offloading the initial filtering of alerts onto an autonomous AI system, since manually performing this task would be inefficient. The described scenario matches exactly the HOTL decision mode of our framework, detailed in Figure \ref{fig:hotl}.

The NeSy AI model takes as input a security alert and produces its priority classification. To properly categorize a security alert, the NeSy AI model leverages information contained in the Shared Knowledge. Common formalisms for representing symbolic knowledge in cybersecurity are Cybersecurity Knowledge Graphs and ontologies, which define security-related concepts and their relationships through subject–predicate–object triples. For example, the Situation and Threat Understanding by Correlating Contextual Observations (STUCCO) ontology \cite{Iannacone2015} provides an intuitive model containing 15 entity types and 115 properties in total, along with entity relations, that represent the cybersecurity domain. STUCCO defines concepts such as software, vulnerabilities, malware, users, and attackers, and supports heterogeneous data sources, including network- and host-based IDSs, thereby enabling network-flow records to be linked with process-level information. The ontology can be used for multiple use cases, including incident response, such as searching through flow records for a specific time slot, and malware detection, e.g., comparing software packages on an impacted host with vulnerability databases.

Particularly in this context, it is important to observe that the Shared Knowledge is not static. New symbolic knowledge may be introduced either by domain experts or through Automatic Knowledge Discovery, for instance, as unknown threats emerge, such as zero-day attacks. Finally, the prioritization decision of the NeSy AI model is fed to the Post hoc delegator, which decides whether to trust the AI output or delegate to the analyst. 

As in the semi-autonomous driving context, neurosymbolic integration also plays a central role in our alert-prioritization framework and offers several potential benefits. By leveraging symbolic contextual information, the NeSy AI model can produce knowledge-guided outputs that are more interpretable to the HOTL and better aligned with up-to-date threat intelligence. Moreover, grounding predictions in trusted, human-readable data sources may improve their reliability and reduce the likelihood of false positives. Neurosymbolic methods also give a suitable basis for implementing the Post hoc delegator. The decision to involve a human expert may depend on the degree to which the NeSy AI model can handle a given input reliably, as reflected in its available symbolic knowledge. Alternatively, delegation may depend on allocation strategies that account for human-related factors, such as the analyst’s expertise or current workload. These conditions can be encoded symbolically and incorporated directly into the delegation policy, granting more reliable control than solely performance-driven solutions.

A concrete instance of an HAIC alert prioritization framework is proposed in \cite{Jalalvand2025}. The authors designed a Learning To Defer From Human Feedback (L2DHF) system in which a predictive AI initially assigns priorities to incoming alerts. The suggested prioritization is compared with the Analyst-Validated Alert Repository (AVAR). If the alert is already known, its assigned priority is overridden by the corresponding value stored in AVAR; otherwise, a Deep Reinforcement Learning (DRL) agent trained with human feedback establishes whether to accept the AI prioritization or invoke the expert. 

Our HOTL formulation can be interpreted as a neurosymbolic extension of L2DHF, thus demonstrating its suitability for alert prioritization. The DRL model, whose reward signal is derived from the analyst’s feedback on alert priority, corresponds to the Post hoc delegator. In L2DHF, the available knowledge is limited to a gold-standard dataset (i.e., AVAR), which is used to look up previously observed alerts, while human feedback is incorporated solely through the reward function. In our vision, NeSy AI enables richer and more effective use of symbolic knowledge and human feedback in this setting by allowing explicit domain knowledge to directly inform both prioritization and delegation.

%% file: section_conclusion.tex
\section{Challenges} \label{sec:concl}
In this section, we analyze the main challenges concerning the realization of the NeSy-HAIC framework. For clarity, we categorize these challenges into two sets related to neurosymbolic integration and collaboration, respectively.
\subsection{NeSy AI-related}
\textit{Knowledge Acquisition.}
A major \say{barrier to entry} for deploying NeSy AI in real-world collaborative applications is the need for extensive domain knowledge, which may be difficult to access and costly or time-consuming to acquire. In general, collecting symbolic knowledge is an expensive process that requires human experts and may be affected by noise and biases. Moreover, knowledge acquisition should not be regarded as a one-time procedure, as symbolic knowledge may evolve throughout the system lifecycle and must be continuously maintained to ensure its completeness and adequacy \cite{Wang2025}. Therefore our NeSy-HAIC framework integrates lifecycle operations for updating symbolic knowledge. To reduce human involvement and optimize knowledge management, particular attention should be devoted to Automatic Knowledge Discovery (AKD), which, in NeSy-HAIC, is performed by the Knowledge Extractor component. Despite the long-standing research on AKD, building routines to extract symbolic knowledge in the desired format, such as knowledge graphs or ontologies, is still a challenge for neurosymbolic systems. \\

\textit{Symbol Grounding Problem.} 
The \say{Symbol Grounding Problem}, described in Section \ref{sec:sgp}, is far from being solved, as the mechanism of translating symbols into computable representations is not unique and scholars do not agree on a single method for realizing neurosymbolic integration. This issue naturally extends to collaborative scenarios, where the most suitable NeSy AI model may vary according to the task type and domain-specific constraints. Some NeSy systems may prioritize interpretability over performance when inspecting internal model behavior is critical for ethical or safety reasons, whereas others may favor predictive performance by reducing the role of the symbolic component in favor of more powerful neural architectures. In collaborative settings where rule enforcement is paramount, approaches based on hard constraints should be preferred over soft constraints incorporated into the loss function, which do not ensure rule satisfaction. The fragmentation of techniques for integrating symbolic and neural systems limits the broader adoption of NeSy AI in HAIC applications, often confining existing solutions to highly customized approaches without a comprehensive view of human-AI interaction. \\

\textit{Scalability and Computational Efficiency.}
The operational efficiency of neurosymbolic systems represents a critical concern, as the integration of symbolic and subsymbolic components often introduces inefficiencies at the hardware execution level and significantly increases latency \cite{Wan2024}. This arises from the divergent computational profiles of the two paradigms: symbolic computing, involving logical operations with sparse vector representations, is typically CPU-intensive and memory-bound, whereas neural networks achieve higher efficiency on GPUs, thereby creating substantial data transfer overhead. In addition, symbolic reasoning often faces scalability challenges, since the increasing expressivity of symbolic formalisms, e.g., the use of complex syntactic structures and universal quantification, can escalate computational complexity \cite{Wang2025}. Grounding symbolic representations into subsymbolic objects may also incur non-negligible computational and time costs, highlighting the need for \textit{ad hoc} strategies to ensure scalability.

\subsection{HAIC-related}

\textit{Human factors and costs.} Human involvement in the decision-making process adds further complexity to collaborative applications. Beyond differences in skill levels, humans are susceptible to fatigue, distraction, and other physical or cognitive conditions that may significantly impair performance. HAIC applications should therefore account for the possibility of human errors during task execution or feedback provision, such as incorrect classification labels that may contain errors or biases. Some approaches to mitigating issues arising from unreliable human input are to aggregate the outputs of multiple human experts, thereby reducing individual variability, as commonly done in crowdsourcing, or to perform quality control checks. Human feedback entails costs associated with training human experts, collecting their responses, and compensating them for their involvement. A crucial challenge in HAIC frameworks lies in minimizing the costs associated with human participation without compromising system accuracy. \\

\textit{Online learning with human and environmental feedback.} One of the main, yet often overlooked, requirements of collaborative systems is to continuously update in response to human inputs, data shifts, and environmental changes. This capability can be achieved only through Online Learning (OL), which entails updating the model every time new data or feedback arrives \cite{Kumar2024}. We argue that OL plays a central role in our NeSy-HAIC framework. In our case, model adaptation includes not only parameter tuning but also the adjustment of symbolic knowledge through human intervention and AKD. We do not delve into the details of how OL should be applied within the NeSy-HAIC framework; however, we acknowledge its importance for the framework’s practical implementation. \\

\textit{Validation.} A thorough assessment of HAIC framework capabilities is challenging due to human involvement in the learning process and the inherent subjectivity of human input \cite{Monarch2021}. The absence of a ground truth for human labels, which can conceal inconsistencies arising from differing interpretations of the observed data, may complicate validation and potentially lead to non-replicable results. This issue, together with the need to repeatedly evaluate the system throughout its lifecycle, calls for the standardization of validation procedures for HAIC frameworks, including NeSy-HAIC.

\section{Conclusion} \label{sec:concl2}
The growing interest in advancing HAIC across multiple application domains is held up by several challenges affecting collaborative systems, including limited interpretability, trust issues, and insufficient data quality. In this paper, we argued that NeSy AI can significantly contribute to the development of better HAIC solutions by establishing a shared symbolic interface between humans and AI models. We reviewed HAIC applications across several domains, including cybersecurity, healthcare, and robotics, identified the main challenges affecting these collaborative systems, and discussed how NeSy AI may help address them. Then, the paper introduced NeSy-HAIC, a general framework for collaborative applications based on neurosymbolic integration, which we also instantiated in two use cases in semi-autonomous driving and alert prioritization. Our work represents an initial attempt to model HAIC from a neurosymbolic perspective and aims to stimulate further research into NeSy AI as a promising paradigm for enhancing human–AI interaction.